\documentclass[11pt]{article}
\pdfoutput=1
\usepackage{jheppub}
\usepackage{amsmath}
\usepackage{amssymb}
\usepackage{dcolumn}
\usepackage{bm}
\usepackage{blkarray}
\usepackage{slashed}                             
\usepackage{hyperref,graphicx}           
\usepackage{url}
\usepackage{caption}
\usepackage{subcaption}
\usepackage{multirow}
\usepackage{units}
\usepackage{physics}
\usepackage{xcolor}
\usepackage{graphicx}
\usepackage{tikz}
\usepackage{siunitx}[=v2]
\usepackage[all]{hypcap}
\usepackage{cleveref}
\usepackage[mathlines]{lineno}
\usepackage{tikz-3dplot}
\usepackage{color}
\usepackage{multirow}
\usepackage{float}

\numberwithin{equation}{section}
\definecolor{darkblue}{rgb}{0,0,0.5}
\definecolor{darkgreen}{rgb}{0,0.5,0}

\newcommand{\cmt}[1]{}

\newcommand{\Lag}[1][]{\mathcal{L}_\text{#1}}

\let\deltafunc\delta
\DeclareDocumentCommand\delta{}{\trigbraces{\deltafunc}}
\usepackage[commandnameprefix=ifneeded]{changes}
\definechangesauthor[name={Yunjia Bao}, color=cyan]{YB}
\definechangesauthor[name={Lingfeng Li}, color=orange]{LL}
\definechangesauthor[name={JiJi Fan}, color=red]{JF}
\setdeletedmarkup{\ifmmode\text{\sout{\ensuremath{#1}}}\else\sout{#1}\fi}

\graphicspath{{./Figure/}}

\begin{document}
\title{\boldmath Electroweak ALP Searches at a Muon Collider }
\author[a]{Yunjia Bao,}
\author[a,b]{JiJi Fan,}
\author[a]{Lingfeng Li}
\affiliation[a]{Department of Physics, Brown University, Providence, RI, 02912, USA}
\affiliation[b]{Brown Theoretical Physics Center, Brown University,
Providence, RI, 02912, U.S.A.}
\emailAdd{yunjia\_bao@brown.edu}
\emailAdd{jiji\_fan@brown.edu}
\emailAdd{lingfeng\_li@brown.edu}

\vspace*{1cm}

\abstract{A high-energy muon collider with center-of-mass energy around and above 10 TeV is also a vector boson fusion (VBF) machine, due to the significant virtual electroweak (EW) gauge boson content of high-energy muon beams. This feature, together with the clean environment, makes it an ideal collider to search for TeV-scale axion-like particles (ALP) coupling to Standard Model EW gauge bosons, which current and other future colliders have limited sensitivities to. We present detailed analyses of heavy ALP searches in both the VBF and associated production channels at a muon collider with different running benchmarks. We also show projected constraints on the ALP couplings in the effective field theory, including an operator with its coefficient not determined by the mixed Peccei-Quinn anomaly. We demonstrate that a muon collider could probe new ALP parameter space and push the sensitivities of the couplings between the ALP and EW gauge bosons by one order of magnitude compared to HL-LHC. The projected limits and search strategies for ALPs could also be applied to other types of resonances coupling to EW gauge bosons. }
\maketitle
\flushbottom

\section{Introduction}
An axion-like particle (ALP) refers to a general periodic pseudo-scalar, $a \cong a + 2\pi f_a$ with $f_a$ the decay constant. ALPs appear ubiquitously in extensions of the Standard Model (SM), usually as pseudo Nambu-Goldstone bosons arising from spontaneous breaking of some global symmetries~\cite{Peccei:1977ur, Peccei:1977hh}. ALPs enjoy a wide range of phenomenological applications\footnote{The earliest examples include the QCD axion from breaking of a global Peccei-Quinn symmetry, which could solve the strong CP problem and is also a cold dark matter candidate~\cite{Peccei:1977ur, Peccei:1977hh, Weinberg:1977ma, Wilczek:1977pj, Kim:1979if, Shifman:1979if, Zhitnitsky:1980tq, Dine:1981rt, Preskill:1982cy, Dine:1982ah, Abbott:1982af}.} and serve as one of the most motivated new physics scenarios. While most of the studies focus on very light ALPs, there has been growing interests to search for a heavy ALP with a mass $m_a$ around or above the weak scale at collider-type experiments. A heavy ALP could acquire its mass from a strongly-interacting hidden sector with a confining scale above the weak scale.\footnote{Ref.~\cite{Gaillard:2018xgk} proposes to use an enlarged but unified color sector to bring the QCD axion mass up to the TeV scale.}
Experimentally, the Large Hadron Collider (LHC) could probe $m_a$ up to the scale of $\sim (2-3)$ TeV, around and above which the parameter space is still widely open. In this article, we will focus on searches of heavy ALPs with $m_a$ of TeV scale, which couple to SM electroweak (EW) gauge bosons, at a possible future muon collider.

Searches for heavy ALP signals could be challenging at LHC and future colliders. A heavy ALP with $m_a \sim \mathcal{O}$(1)~TeV requires highly energetic beams for direct production. In addition, its one-loop-suppressed couplings and a large $f_a$ further restrain the signal rate. Consequently, combining a high integrated luminosity and low background level is necessary for detecting it. A variety of direct heavy ALP searches are available (\textit{e.g.}, \cite{Jaeckel:2012yz, Mimasu:2014nea,Jaeckel:2015jla,Knapen:2016moh,Bauer:2017ris,Craig:2018kne,Lee:2018pag,Bauer:2018uxu, Hook:2019qoh,Carmona:2021seb,Wang:2021uyb,Agrawal:2021dbo, Florez:2021zoo,Liu:2021lan}), constraining visible ALPs at LEP and LHC. Future circular $e^+e^-$ colliders~\cite{FCC:2018evy,CEPCStudyGroup:2018ghi} are not optimal for the detection of TeV-scale ALPs as their achievable energies are limited by the synchrotron radiation energy loss. Future linear $e^+e^-$ colliders~\cite{Bambade:2019fyw,CLICdp:2018cto}, although almost free from radiation energy loss, suffer from smaller luminosities as each bunch is dumped instead of stored after the first crossing. Heavy ALPs could be efficiently produced at future hadron colliders~\cite{FCC:2018vvp}. However, the signal could easily be overwhelmed by the enormous QCD backgrounds or vulnerable to large systematic uncertainties.

A future muon collider provides the potential solution to all the issues regarding energy, luminosity, and background cleanliness~\cite{Skrinsky:1981ht,Neuffer:1983xya,Neuffer:1986dg,Barger:1995hr,Barger:1996jm,Ankenbrandt:1999cta}.  As $m_\mu \approx 207 m_e$, the synchrotron radiation energy loss is under control since the energy loss rate is proportional to $m_{\ell}^{-4}$. The circular collider design with a high luminosity could be achievable even for multi-TeV muon beams. From the conservative projection~\cite{Snowmass,Aime:2022flm}, the expected integrated luminosity reaches~$\mathcal{O}(1)$~ab$^{-1}$ for TeV-scale muon colliders and could potentially increase further. Muon also carries a significant portion of beam energy as an elementary particle. In contrast, multiple quarks and gluons share the beam energy at a future hadron collider, with the high energy tip of parton distribution functions (PDFs) suppressed. When searching for a heavy ALP that couples to EW vector bosons, the advantage of a muon collider is even more explicit since the beam muon, unlike a beam proton, radiates a significant fraction of its energy to all EW vector bosons. For example, about $5\%$ of the total beam energy is carried by $\gamma,~W$, and $Z$ in a TeV-scale muon beam~\cite{Han:2020uid}, while this fraction drops to below $1\%$ in a proton beam~\cite{Bertone:2017bme}. In addition, a muon collider can directly produce an ALP almost as heavy as the center-of-mass energy $\sqrt{s}$ from $\mu^+\mu^-$ annihilations in association with an EW vector boson, which is unlikely at hadron colliders. Last but not least, the inclusive event rate of muon collisions is much lower than that of hadronic collisions. It is well known that the proton-proton inclusive cross section scales as the proton size ($\sim \Lambda_{\rm QCD}^{-2}\sim\mathcal{O}$(mb)). On the other hand, muons are elementary particles, and their inclusive cross section remains small at large $\sqrt{s}$, dominated by vector boson fusion (VBF) processes. This feature greatly reduces the overall background level and background-induced systematic uncertainties.

The main difficulty of the muon collider concept comes from the short muon lifetime $\sim 2~\mu$s. It is highly challenging to produce low emittance muon beams and store them in the collider ring before they decay. Thanks to the progress of accelerator physics, the particle physics community's interest in a high-energy muon collider grows again. Several leading approaches have been developed quickly over the past years~\cite{Palmer:2014nza,Delahaye:2019omf}. The US Muon Accelerator Program (MAP)~\cite{Delahaye:2013jla,Delahaye:2014vvd,Ryne:IPAC2015-WEPWA057,Long:2020wfp} focuses on proton-driven muon sources where secondary pions decay to muons. The program's target scenario covers a wide range of beam energy, from $\mathcal{O}$(1-10)~GeV neutrino factories to Higgs factories and multi-TeV colliders. Such proton-driven muon beams occupy a large phase space volume, which need to be cooled significantly before muons decay. The issue is addressed by the development of Muon Ionization Cooling Experiment (MICE)~\cite{Mohayai:2018rxn,Blackmore:2018mfr,MICE:2019jkl}, which demonstrates the potential of cooling muon beams over short time scales. More recently, the novel approach of Low Emittance Muon Accelerator (LEMMA) allows high luminosity muon beams to be produced with low transverse emittance~\cite{Antonelli:2015nla,Long:2020wfp}. The design uses a $45$-GeV positron beam to produce muon pairs at the threshold, resulting in a small beam transverse emittance without extra beam cooling. The technology may allow a muon collider operating at even higher energies up to $\sqrt{s}=\mathcal{O}(100)$~TeV but with limited luminosities.

The abundant VBF interaction rates and low SM backgrounds make a high energy muon collider optimal for precision SM tests and beyond-the-Standard-Model (BSM) searches. With high energy and high luminosity muon beams, a future muon collider enables a plethora of possible discoveries. Latest phenomenological studies demonstrate the advantages of a muon collider in many different contexts. They already cover quite a few interesting aspects, such as measuring various Higgs properties~\cite{Chiesa:2020awd,Han:2020pif,Chiesa:2021qpr,Cepeda:2021rql,Chen:2021pqi,Buonincontri:2022ylv}, probing dark matter models~\cite{Han:2020uak,Capdevilla:2021fmj,Medina:2021ram}, investigating models motivated by flavor physics~\cite{Huang:2021biu,Asadi:2021gah,Haghighat:2021djz,Bandyopadhyay:2021pld} or the $g_\mu -2$ anomaly~\cite{Capdevilla:2020qel,Buttazzo:2020ibd,Yin:2020afe,Capdevilla:2021rwo,Chen:2021rnl,Li:2021lnz,Dermisek:2021mhi,Capdevilla:2021kcf}, higher-dimensional operator analyses~\cite{DiLuzio:2018jwd,Buttazzo:2020uzc,Spor:2022mxl,Chen:2022msz}, and other BSM studies~\cite{Eichten:2013ckl,Chakrabarty:2014pja,Buttazzo:2018qqp,Bandyopadhyay:2020otm,Liu:2021jyc,Han:2021udl,Sen:2021fha,Liu:2021akf,Cesarotti:2022ttv,Bandyopadhyay:2020mnp}. General prospects of muon collider phenomenology can be found in~\cite{Costantini:2020stv,AlAli:2021let,Franceschini:2021aqd}. However, the panorama of future muon collider phenomenology remains to be fully explored.

Given all the potential advantages outlined above, we implement and report in this paper a detailed analysis of TeV-scale ALPs coupling to EW gauge bosons at a future muon collider. We show that the search of electroweakly coupled ALPs provides another showcase for the great physics potential of a high-energy muon collider. The search strategies and projected sensitivities on the cross sections could be applied to other types of EW resonances.

The paper is organized as follows. In Section~\ref{sec:basic}, we present the effective field theory (EFT) of an ALP that couples to SM EW gauge bosons. In Section~\ref{sec:HLLHC}, we present the common production mechanisms of heavy ALPs at a collider. The HL-LHC projections on ALP couplings are deduced from current LHC limits, as HL-LHC is the inevitable step en route to any future colliders. In Section~\ref{sec:Simulation}, details of simulations for ALP searches at a muon collider are presented. Section~\ref{sec:VBF} enumerates seven analyses on ALPs produced in the VBF channel, aiming at seven final states. Meanwhile, we show that the analysis results could apply to general ALP couplings and other new physics resonance models. The gain in sensitivity from forward-region information is also discussed. Another important ALP production mechanism, associated production with the tri-photon final state, is analyzed in Section~\ref{sec:Va}. In Section~\ref{sec:EFT}, the projected constraints on the EFT couplings are deduced from limits derived in the previous sections. We conclude in Section~\ref{sec:Conclusion}.

\section{The Effective Theory of ALP coupling to EW Gauge Bosons}
\label{sec:basic}

We will focus on the EFT of ALP coupling to the EW gauge bosons in the SM. Here, the heavy axion of interest has a mass of ${\cal O}$(TeV). Before electroweak symmetry breaking (EWSB), the Lagrangian reads:
\begin{align}
	\Lag &= \frac{1}{2} \qty[\qty(\partial a)^2 - m_a^2 a^2] 
	+ \qty(\frac{g_1}{4\pi})^2 C_{BB} \frac{a}{f_a} B_{\mu\nu} \widetilde{B}^{\mu\nu} \nonumber\\
	& + \qty(\frac{g_2}{4\pi})^2 C_{WW} \frac{a}{f_a} W_{\mu\nu}^i \widetilde{W}^{i;\mu\nu} 
	+ \qty(\frac{g_1}{4\pi}) \qty(\frac{g_2}{4\pi}) C_{BW} \frac{a}{f_a} B_{\mu\nu} \widetilde{W}^{3;\mu\nu} \, ,
	\label{eq:EFT1}
\end{align}
where $W(B)_{\mu\nu}$ are the field strength tensors of SM $SU(2)_W(U(1)_Y)$ gauge groups before EWSB, $\widetilde{W}(\widetilde{B})_{\mu\nu}\equiv \epsilon_{\rho\sigma \mu\nu} W(B)^{\rho\sigma}/2$ are the corresponding dual field strengths, $i=1, 2, 3$ are the $SU(2)_W$ indices, and $g_{1(2)}$ are the corresponding coupling constants of $U(1)_Y (SU(2)_W)$. These effective couplings could be generated by integrating out heavy fermions carrying EW charges at one loop. The coefficients are thus suppressed by the loop factor, $1/(4\pi)^2$.\footnote{The loop factors here are expected in general so that we could reproduce the axion coupling to photons with the right form, \textit{i.e.}, $\alpha/(4\pi)$ times a quantized coefficient with $\alpha$ the fine structure constant, after EWSB.}
Note that other conventions for the EFT Lagrangian exist. In particular, another widely used convention is to rescale the coefficients $\{C_{WW}, C_{BB}, C_{BW}\} \to \{ (g_2/4\pi)^2 C_{WW}, (g_1/4\pi)^2 C_{BB}, \\ g_1 g_2/(4\pi)^2 C_{BW}\}$ to absorb the gauge couplings and loop factors~\cite{Brivio2017}. We will follow the convention in Eq.~\eqref{eq:EFT1} throughout the analysis, while the other convention will also be presented in some plots for the readers' convenience. We will ignore other possible ALP couplings like the ones to SM fermions and the Higgs in our study since these couplings could lead to different production channels and detection strategies beyond the scope of this paper, \textit{e.g.}, $s$-channel muon annihilation $\mu^+\mu^- \to a(+\gamma) \to \mu^+\mu^-(+\gamma)$ if an axion has a sizable coupling to muon.

It is noteworthy that a non-zero $C_{BW}$ coupling in Eq.~\eqref{eq:EFT1} is not always included in the literature. An ALP coupling is usually proportional to associated global anomaly coefficient, which is true for an ALP coupling to massless SM gauge bosons, such as photons or gluons. Yet as discussed in Refs.~\cite{Alonso-Alvarez:2018irt,BlindToAnomaly,Bonnefoy2020,Quevillon:2021sfz,Bonilla:2021ufe}, this coupling could arise in several ALP models and shall be considered in a generic analysis. Although there is no non-vanishing $U(1)_{\text{global}} \times U(1)_Y \times SU(2)_W$ anomaly coefficient for $C_{BW}$, the coupling term turns out to arise from dimension-7 operators such as $a H^\dagger \vb*{\tau}^i H W^i_{\mu\nu} \tilde{B}^{\mu\nu}/\Lambda^3$, where $H$ is the SM Higgs doublet, $\vb*{\tau}^i$'s are $SU(2)_W$ generators, and $\Lambda$ is the UV cutoff scale. After EWSB, this results in an operator of the form $\qty(a/\Lambda) \qty(v_\text{EW} / \Lambda)^2 \tilde{B}^{\mu\nu} W_{\mu\nu}^3$, in which $v_\text{EW}$ is the Higgs VEV. This term could be numerically important when $\Lambda$ is not far beyond the EW scale. There have been efforts to construct models in which such terms appear. In the standard DFSZ axion model~\cite{Dine:1981rt,Zhitnitsky:1980tq}, the light axion is a linear combination of angular modes of the two Higgs doublets and the PQ scalar~\cite{Srednicki1985, Buen-Abad2021}. The top loop generates a non-zero $C_{BW}$ coupling~\cite{Bonilla:2021ufe}. Another example of a possible UV completion is constructed in Ref.~\cite{Bonnefoy2020} via a DFSZ-type extension in which heavy EW charged chiral fermions are integrated out. In this scenario, the new fermions receive their masses mainly from EWSB and are thus within reach of current and near-future collider searches. One may also note that other operators exist beyond dimension 7; however, at and below dimension 7, the operators listed in Eq.~\eqref{eq:EFT1} are equivalent to the complete set of gauge-invariant operators that involve one axion field and EW gauge bosons.  In this paper, we remain agnostic about the UV completion of the $C_{BW}$ term in the ALP EFT.

After EWSB, the ALP couplings to the gauge boson mass eigenstates can be written as:
\begin{equation}
	\Lag \supset \frac{1}{2} \qty(\partial a)^2 - \frac{1}{2} m_a^2 a^2 + \frac{\alpha}{4\pi} \frac{a}{f_a} \bigg( C_{\gamma\gamma} F \widetilde{F} + 2 C_{\gamma Z} F \widetilde{Z} + C_{ZZ} Z \widetilde{Z} + \frac{2}{s_W^2} C_{WW} W^+ \widetilde{W}^-\bigg) \, ,
	\label{eq:EFT2}
\end{equation}
with
\begin{equation}
	F \widetilde{Z} 
	\equiv \frac{1}{2} \epsilon^{\mu\nu\rho\sigma} F_{\mu\nu} Z_{\rho\sigma}~, ~s_W \equiv \sin \theta_W~,~c_W \equiv \cos \theta_W~,
\end{equation}
where $F,~Z,~W$ are EW field strengths, $\alpha$ is the fine structure constant, and $\theta_W$ is the Weinberg angle. The magnitude $C_{WW}$ remains the same as in Eq.~\eqref{eq:EFT1}. The other three effective couplings after EWSB are related to the linear combinations of the original ones as
\begin{equation}
	\begin{gathered}
		C_{\gamma\gamma} \equiv C_{BB} + C_{WW} + C_{BW} \, , \quad 
		C_{\gamma Z} \equiv  C_{WW} \frac{c_W}{s_W}  - C_{BB} \frac{s_W}{c_W} + \frac{1}{2} C_{BW} \qty(\frac{c_W}{s_W} - \frac{s_W}{c_W}) \, , \\
		C_{ZZ} \equiv C_{WW} \frac{c_W^2}{s_W^2} + C_{BB} \frac{s_W^2}{c_W^2} - C_{BW} \, .
	\end{gathered}
	\label{eq:couplingsEWSB}
\end{equation}
With the couplings above, the partial decay widths of axion to vector bosons are given by 
\begin{equation}
	\begin{gathered}
		\Gamma(a \to \gamma \gamma) = \frac{1}{2}  \qty(\frac{C_{\gamma\gamma}}{f_a})^2 \frac{\alpha^2 m_a^3}{32 \pi^3} \, ,  \\
		\Gamma(a \to \gamma Z) = \qty(\frac{C_{\gamma Z}}{f_a})^2 \frac{\alpha^2 m_a^3}{32 \pi^3} \qty[1 - \qty(\frac{m_Z}{m_a})^2]^3 \, ,   \\
		\Gamma(a \to ZZ) = \frac{1}{2} \qty(\frac{C_{Z Z}}{f_a})^2 \frac{\alpha^2 m_a^3}{32 \pi^3} \qty[ 1 - \qty(\frac{2m_Z}{m_a})^2 ]^{3/2} \, ,  \\
		\Gamma(a \to W^+W^-) = \qty(\frac{C_{WW}}{f_a})^2 \frac{\alpha^2 m_a^3}{32 s_W^4 \pi^3} \qty[ 1 - \qty(\frac{2m_W}{m_a})^2 ]^{3/2} \, ,
	\end{gathered}
	\label{eq:decay}
\end{equation}
where $m_Z$ and $m_W$ are the masses of the $Z$ and $W$ gauge bosons.

\section{ALP Collider Signals and (HL)-LHC Constraints}
\label{sec:HLLHC}

In this section, we will introduce two production channels of heavy ALPs based on the EFT in Eq.~\eqref{eq:EFT1} and discuss projected constraints on the ALP couplings to EW gauge bosons at HL-LHC. We will also comment on the indirect probes where the heavy ALP is a mediator. 

\subsection{Vector Boson Fusion}
\label{ssec:VBF}

\begin{figure}[h]
\centering
\includegraphics[width=7 cm]{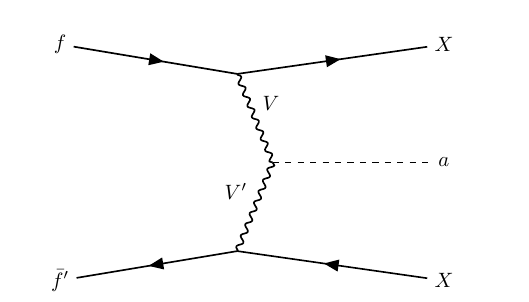}
	\caption{Feynman diagram for the VBF production of ALP at a collider. }
	\label{fig:VBF_Cartoon}
\end{figure}

VBF channels refer to both $WW$ fusion and neutral boson fusion ($ZZ$, $Z\gamma$, and $\gamma\gamma$ fusion) processes, as shown in Fig.~\ref{fig:VBF_Cartoon}. They could be schematically written as
\begin{equation}
	f\bar{f}^{\prime} \to VV^\prime+X{\text {'s}} \to a +X{\text {'s}} \, ,
\end{equation} 
where $f$ and $f^\prime$ are beam fermions, $V$ and $V^\prime$ are SM vector bosons, and $X$'s denotes recoiling beam remnants. The representation above applies to both hadron and lepton colliders with different choices of $f^{(\prime)}$ and corresponding $X$'s. Due to the enhanced amplitude when $X$'s are collinear with the beams, $X$'s are mainly present in the forward regions with large $|\eta|$'s, which could be helpful to veto non-VBF backgrounds. However, the detector performance is usually limited in these regions. The VBF-produced ALPs will also have low $p_T$'s on average and generate a sharp invariant mass peak $\approx m_a$ if their final states are fully visible. Since the width $\Gamma_a\ll m_a$ for realistic couplings, it is justified to take the narrow-width approximation and make the ALP production and decay independent to each other. 

At hadron colliders like the LHC, the initiating fermions $f^{(\prime)}$ are quarks and the recoiling remnant $X$'s contain two (or even more) jets in the high-$|\eta|$ region. LHC has already implemented several relevant searches with the $\sqrt{s}=$13~TeV dataset, $e.g.$, searches for diphoton resonances~\cite{CMS:2016kgr,ATLAS:2021uiz}, inclusive $WW$ or $ZZ$ resonances~\cite{ATLAS:2018sxj,ATLAS:2018sbw,CMS:2019qem,ATLAS:2020fry}, and $Z\gamma$ resonances~\cite{ATLAS:2016mti,CMS:2017dyb}. The large SM backgrounds from hadronic interactions significantly hinder signal reconstruction and affect the discovery potential. For example, soft particles from pileup could pose serious challenges to reconstruct mother particles in the decay chains, especially for hadronic $W$ or $Z$ fat jets as their large jet areas contain more pileup particles~\cite{Cacciari:2008gn}. Most exclusion limits at 95\% C.L. on $\sigma(pp\to a +X\text{'s})\times \text{BR}(a\to VV^\prime)$ are of $\mathcal{O}(1-10)$~fb when $m_a\approx 1$~TeV. These limits strengthen by about one order of magnitude when $m_a\approx 2$~TeV since the SM backgrounds drop rapidly as the hard scattering energy scale increases.

\begin{figure}[h]
	\centering
	\includegraphics[width=8.5 cm]{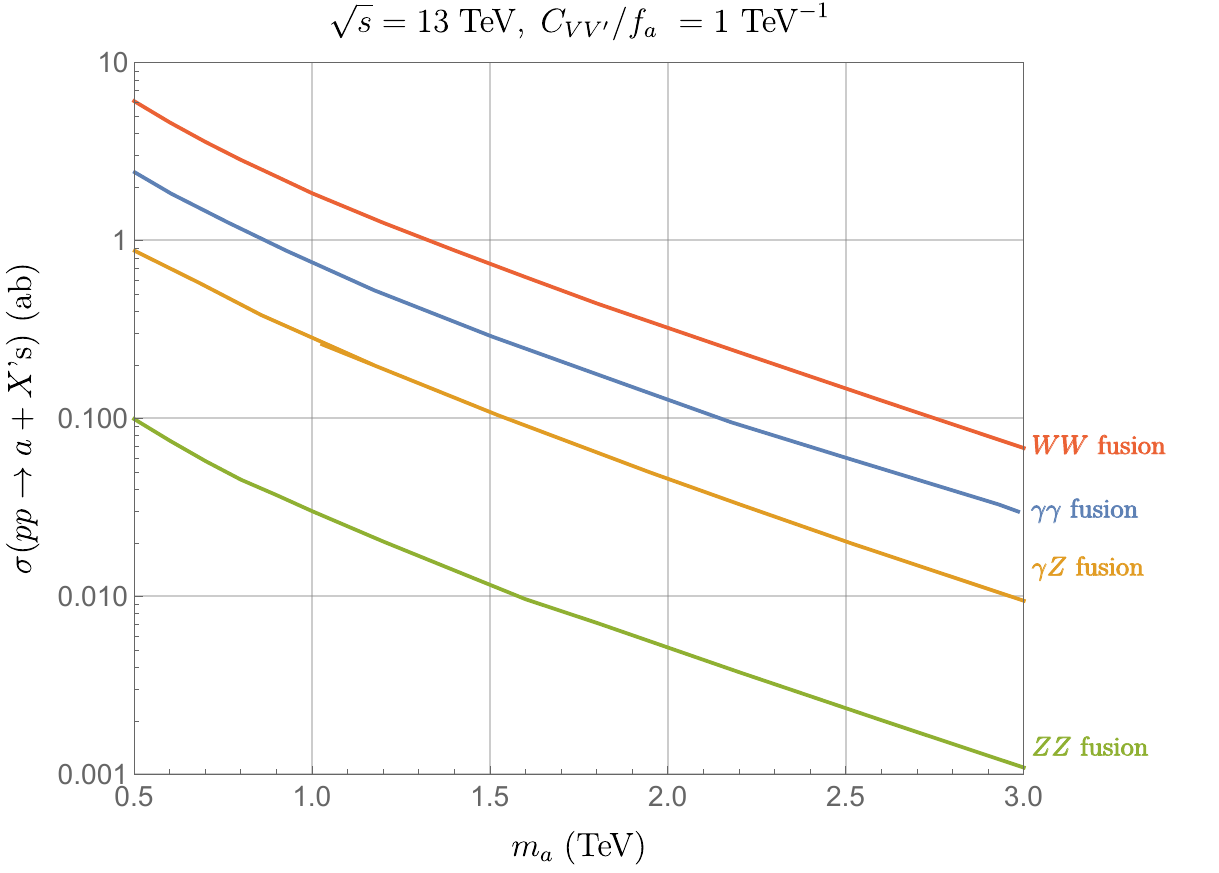} \hspace{-0.5 cm}
	\includegraphics[width=7cm]{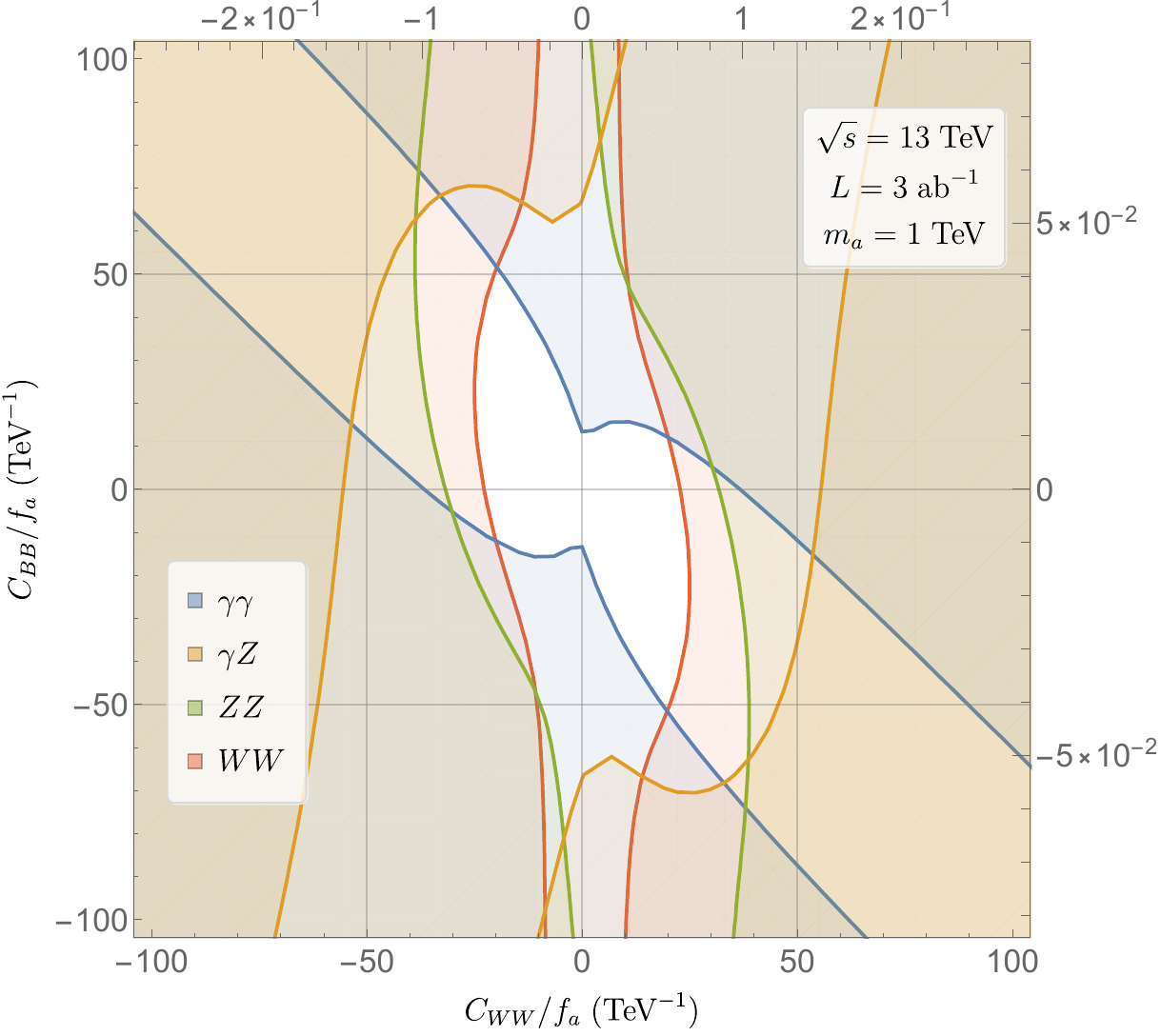}
	\caption{\textbf{Left:} Different VBF production rates as functions of $m_a$ at the 13-TeV LHC, fixing $C_{VV^\prime}/f_a$ = 1 TeV$^{-1}$. It should be noted that it is impossible to obtain all $C_{VV^\prime}/f_a$ = 1 TeV$^{-1}$ simultaneously since there are only three Wilson coefficients, $(C_{BB}/f_a, C_{BW}/f_a, C_{WW}/f_a)$, before EWSB. Nonetheless, treating $\gamma$, $Z$, and $W$ as partons of protons allows us to show four fusion channels individually as benchmarks. \textbf{Right:} Projected HL-LHC $95\%$ C.L. constraints of $C_{WW}/f_a$ and $C_{BB}/f_a$, fixing $C_{BW} =0$ and $m_a = 1$ TeV. The bottom and left axes indicate values of the axion couplings defined in Eq.~\eqref{eq:EFT1}, while the top and right axes indicate another popular EFT convention in~\cite{Brivio2017}. }
	\label{fig:VBFpheno}
\end{figure}

We adopt the four VBF rates of ALP productions from Ref.~\cite{Molinaro:2017rpe}, using the LUXqed photon PDF~\cite{Manohar:2016nzj} and matrix element (ME) method. Notice that the interference between initial state $\gamma$ and transverse $Z$ bosons is ignored as the PDF set including $W$ and $Z$ bosons is not yet available. In addition, such an interference is unlikely to change the ALP production rate significantly~\cite{Fornal:2018znf,Buarque:2021dji}. Hence, we leave a more precise determination of ALP production rates at hadron colliders to future work. In the left panel of Fig.~\ref{fig:VBFpheno}, we show the corresponding production cross section $\sigma(pp\to a+X{\text {'s}})_{\rm VBF}$ as a function of axion mass $m_a$, normalizing each $C_{VV^\prime}/f_a = 1$ TeV$^{-1}$ after EWSB. Hierarchies between processes originate from vector boson properties and definitions in Eq.~\eqref{eq:EFT2}. For example, the $WW$ fusion rate is enhanced by the large $ s_W^{-2}$ factor in Eq.~\eqref{eq:EFT2} and becomes much larger than the other three. It is followed by the $\gamma\gamma$ fusion rate as it is easy for the energetic proton beam to radiate a massless photon. For $m_a, f_a \gtrsim 1$ TeV, and assuming ${\cal O}(1)$ EFT couplings, the typical production rates are $\lesssim 10^{-2}$~fb, which are at least two orders of magnitudes below the current LHC limits.

To estimate the HL-LHC sensitivities on couplings between ALP and EW gauge bosons, we take the current limits and scale them with the integrated luminosity. We assume that systematic uncertainties will scale as statistical ones.\footnote{We also assume difficulties induced by the high pileup level in the high luminosity era~\cite{CMS:2013xfa} can be mitigated by techniques such as time-of-flight measurements~\cite{Butler:2019rpu} and machine learning~\cite{Komiske:2017ubm,ArjonaMartinez:2018eah}.} The constraints on $C_{WW}/f_a$ and $C_{BB}/f_a$ are plotted in the right panel of Fig.~\ref{fig:VBFpheno} for a 1-TeV ALP, fixing $C_{BW}=0$. We take into account of $a\to W^+W^-$ and $a \to ZZ$ constraints from~\cite{ATLAS:2020fry}, $a\to Z\gamma$ constraint from the scalar production benchmark in Ref.~\cite{ATLAS:2018sxj}, and $a\to \gamma\gamma$ constraint from~\cite{ATLAS:2021uiz}.\footnote{We ignore the signal efficiency differences between ALP and the benchmark models ($e.g.,$ KK graviton or radion) used in the LHC analyses.} For $f_a \gtrsim m_a \sim$ TeV, HL-LHC could only probe $C_{VV^\prime}$ of ${\cal O}(10)$ or higher. The situation still persists when $C_{BW}\neq 0$. In another convention of ALP EFT~\cite{Brivio2017}, where the loop factors and gauge couplings in Eq.~\eqref{eq:EFT1} are absorbed in the definition of couplings, the projected sensitivities will appear stronger, as shown by the top and right axes of the right panel in Fig.~\ref{fig:VBFpheno}.

\subsection{Associated Production}
\label{ssec:associated}
\begin{figure}[h!]
\centering
\includegraphics[width=6.5 cm]{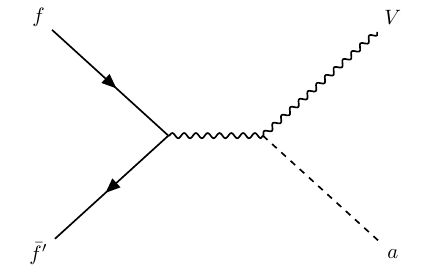}
	\caption{Feynman diagram for ALP production associated with a SM vector boson.}
	\label{fig:Asso_Cartoon}
\end{figure}
Another important ALP production mechanism is the associated production of an ALP with an EW vector boson, sometimes also called ALP-strahlung:
\begin{equation}
	f\bar{f}^\prime \to Va,~V = \gamma,~W^\pm,~Z \, .
\end{equation}
As the ALP further decays to two bosons, the typical collider signal is a tri-boson final state, $VV^\prime V^{\prime\prime}$, with $V^\prime V^{\prime\prime}$ forming a narrow resonance. The powers of loop factors and $f_a$ suppressions in these processes are the same as the VBF production channels. Since we ignore the ALP couplings to SM fermions, the associated production at colliders is dominated by the $s$-channel diagram through an off-shell vector boson, as shown in Fig.~\ref{fig:Asso_Cartoon}.

\begin{figure}[h!]
	\begin{center}
		\includegraphics[width=8.5cm]{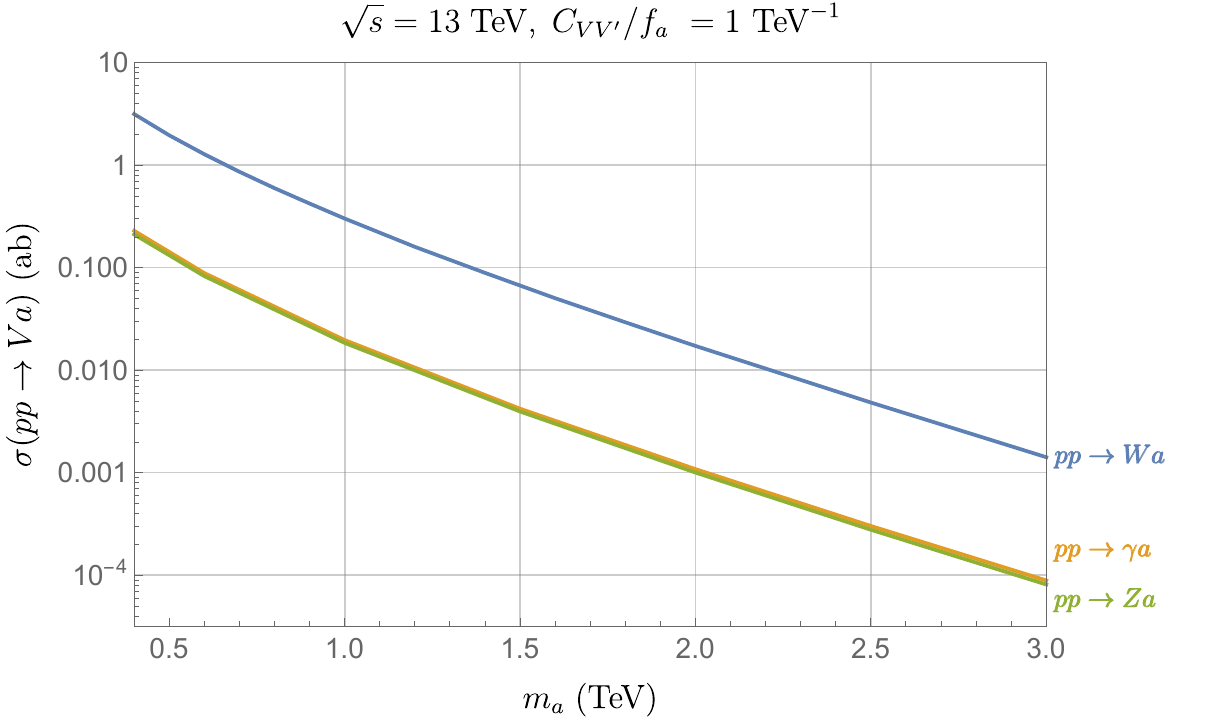}
	\end{center}
	\caption{Different associated production rates as a function of $m_a$ at the 13-TeV LHC, fixing $C_{VV^\prime}/f_a$ = 1 TeV$^{-1}$.
	}
	\label{fig:3V}
\end{figure}

At LHC, several searches for tri-boson signals are implemented with the 13-TeV data. There are limits on cross sections of $W\gamma\gamma$ and $Z\gamma\gamma$ processes~\cite{CMS:2021jji}. Measurements of $WWW$, $WWZ$, $WZZ$ and $ZZZ$ final states are also presented in~\cite{ATLAS:2019dny,CMS:2020hjs,CMS:2019mpq}. Nonetheless, for $WZ\gamma$ and $WW\gamma$ final states, the 13-TeV analysis is not available yet, leaving the latest constraints from the 8-TeV data~\cite{ATLAS:2017bon}. Most analyses above are non-resonant measurements of tri-boson rates, which all turn out to be compatible with the SM predictions. The typical uncertainties on $\sigma(VV^\prime V^{\prime\prime})$ are of $\mathcal{O}(1-100)$~fb depending on the final states. The resonance $Wa(\to WW)$ limit is also deduced in Ref.~\cite{CMS:2019mpq}, covering $m_a\in [200,600]$~GeV range.\footnote{The signal regions, however, do not target narrow $WW$ resonances; thus, the limits are still qualitatively similar to non-resonant ones.} Fig.~\ref{fig:3V} shows the typical $pp \to Va$ cross sections at $\sqrt{s}=13$~TeV, with each relevent $C_{VV^\prime}/f_a=1$~TeV$^{-1}$ after EWSB. The production rates drop faster than the VBF ones as $m_a$ increases, due to the power suppression of energy from the $s$-channel propagator.

Because of the lower rates and essentially non-resonant limits, the projected HL-LHC bounds from $Va$ associated production are even weaker than the VBF ones in the right panel of Fig.~\ref{fig:VBFpheno}. For a 1-TeV ALP (with $f_a$ of the same order), typical constraints on $C_{VV^\prime}$ are $\gtrsim \mathcal{O}(10^2)$, dominated by the $W\gamma\gamma$ and $Z\gamma\gamma$ measurements with lower experimental uncertainties. Even for future resonance searches associated with an extra boson, the small $Va$ production rates still make this channel less relevant at HL-LHC.

\subsection{Non-resonant (Indirect) Signals}
\label{ssec:indirect}
Inevitably, heavy ALPs coupled to the SM could also modify event rates as a mediator. In this case, there will be no narrow ALP resonances in the final states. Since the most significant correction stems from the interference terms between the SM diagrams and ALP-exchanging diagrams, the modifications have the same loop factor and $f_a$ suppressions as the VBF processes. However, as discussed in the last section, non-resonant searches have large SM backgrounds and are subject to significant systematic uncertainties. The current LHC data from this channel are only relevant when $m_a\lesssim m_Z$~\cite{Bonilla:2022pxu}.

For $m_a \gg m_Z$, the impact of virtual ALPs can be described by several dimension-8 anomalous quartic gauge coupling (aQGC) operators. They are suppressed by powers of $m_a^{-1}$, in addition to $f_a^{-1}$ and loop factors from the EFT in Eq.~\eqref{eq:EFT1}. The natural scaling of aQGC terms from ALP mediation follows
\begin{equation}
	\mathcal{L}_{\rm eff} \supset \frac{C_{VV^\prime}^2\alpha^2}{16\pi^2 f_a^2 m_a^2}\mathcal{O}_{\rm SM}^8 \, ,
\end{equation}
where $\mathcal{O}_{\rm SM}^8$ represents dimension-8 aQGC operators. The limits can be set by both multi-boson~\cite{Green:2016trm} and vector boson scattering (VBS)~\cite{Buarque:2021dji} measurements. The extra suppression factor $\sim \alpha^2/(16\pi^2)\lesssim 10^{-6}$ strongly hinders the discovery potential of this non-resonance method. According to the HL-LHC performance study~\cite{Azzi:2019yne}, the coefficient of the dimension-8 operators above can be measured down to $\sim 0.5~\text{TeV}^{-4}$ in the high luminosity phase. To make the signal detectable for $m_a, f_a \simeq $1~TeV, $C_{VV^\prime}$'s need to be as large as $\mathcal{O}(10^3)$. Such bounds are much weaker than those given by the direct searches. Though corrections to the discussions above may arise from finite $m_a\lesssim \sqrt{s}$, it is safe to conclude that the non-resonant/indirect approach will be unlikely to play an important role in heavy ALP searches at the HL-LHC.

\section{Muon Collider Phenomenology: Simulation Setups}
\label{sec:Simulation}

In the forthcoming sections, we will study the potential of searching a heavy ALP, in particular, through its coupling to EW gauge bosons at a future muon collider. 

Throughout this work, we use \textsc{MadGraph 5}~\cite{Alwall:2014hca} to generate parton-level hard processes for both ALP signals and SM backgrounds. We then use \textsc{Pyhthia 8}~\cite{Sjostrand:2007gs} to handle hadronic and electromagnetic shower effects. The detector effects are simulated by \textsc{Delphes 3}~\cite{deFavereau:2013fsa} with its built-in muon collider detector template. 

For simplicity, we follow the detector template's default parameters for reconstructing elementary objects, $e.g.$, photon $\gamma$ and light leptons $\ell$. A photon with transverse momentum $p_T>0.5$~GeV is considered isolated if the scalar sum of $p_T$ over all the other particles inside a cone with a radius $\Delta R=0.1$ around the photon is smaller than 20\% of the photon's $p_T$. For simplicity, we adopt the approximation that the detector efficiencies and resolutions of electrons and muons are the same. In practice, only processes with muonic final states are simulated, with their event rates rescaled to the inclusive ones of both electrons and muons accordingly. The $p_T$ and isolation requirements for isolated muons are the same as the photon ones. Both isolated muons and photons must satisfy $|\eta| < 2.5$. Besides, muons in VBF beam remnants encode critical information like the identities of initial-state vector bosons. The ability to detect beam muons, usually in the forward region, enables a better signal background discrimination~\cite{Han:2020uak}. These recoiling muons are then measured with the default forward muon spectrometer in the muon collider template, covering the $2.5<|\eta|<8$ region. We use the Valencia algorithm to cluster jets for hadronic ALP decay final states, which was reported to provide better performances at high-energy lepton colliders~\cite{Boronat:2014hva}. The $W$ and $Z$ bosons from ALP decays are highly boosted on average. Therefore, we cluster them into fat jets with cone sizes $R=1.0$ or $1.2$, depending on the detection channels. All fat jets must have $p_T > 200$~GeV, above which $W/Z$ bosons are sufficiently boosted. We also assume the tracker's resolution on the impact parameter is much smaller than $\tau$'s lifetime $\sim 87~\mu$m. Charged particles from $\tau$ decays are thus long-lived enough to have significant impact parameters. Such features are distinctive from prompt ALP decays. Therefore, we safely ignore any backgrounds having $Z\to\tau\tau$ decays.

We follow the muon collider's beam energy and integrated luminosity benchmarks in~\cite{AlAli:2021let,Snowmass,Aime:2022flm}. For each beam energy, the analysis results are projected according to the conservative ($L_{\rm con}$) or the optimistic ($L_{\rm opt}$) scenario of the integrated luminosity. In particular, we choose four different energy benchmarks and corresponding luminosities:
\begin{equation}
	\begin{array}{r c *{3}{l<{,}} l l}
		\sqrt{s} & = & 10 & 14 & 30 & 50 & \si{\TeV} \, ,\\
		L_{\rm con} & = & 10 & 10 & 10 & 10 & \si{\per\atto\barn}\, ,\\
		L_{\rm opt} & = & 10 & 20 & 90 & 250 & \si{\per\atto\barn} \, .
	\end{array}
	\label{eqn:collider_benchmarks}
\end{equation}

\subsection{Simulating ALP Signals}
\label{ssec:signals}
\begin{figure}[h!]
	\centering
	\includegraphics[width=0.6 \textwidth]{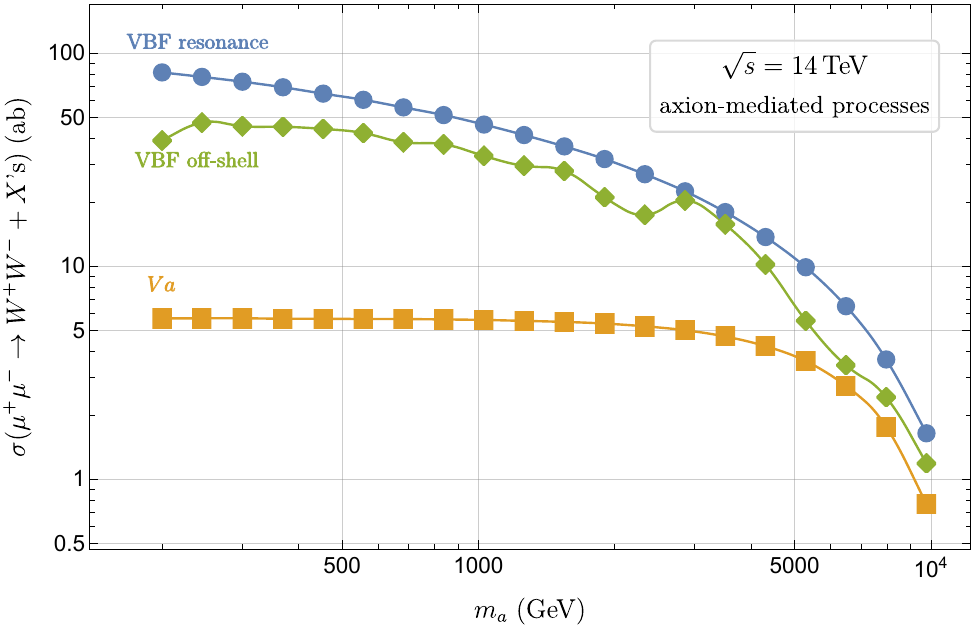}
	\caption{Benchmark production rates of the $W^+W^-$ final state via the heavy axion as a function of its mass at a future muon collider with $\sqrt{s} = 14$ TeV. The $Va$ channel includes both $\gamma a$ and $Z a$ processes with $a\to W^+W^-$ decay. The indirect effect of an off-shell exchange of an axion is estimated by $\sigma_{\text{off-shell}} \equiv \sigma({\mu^+\mu^-}\to W^+W^-+X\text{'s})|_{\text{off-shell}+{\rm SM}}-\sigma({\mu^+\mu^-}\to W^+W^-+X\text{'s})|_{\rm SM}$, in which the $\sigma({\mu^+\mu^-}\to W^+W^-+X\text{'s})|_{\rm SM}$ is the SM VBS cross section, and $\sigma({\mu^+\mu^-}\to W^+W^-+X\text{'s})|_{\text{off-shell}+{\rm SM}}$ is the cross section of the same process including $t$-channel and off-shell $s$-channel ALP diagrams. Similarly, the corresponding VBF resonance cross section is also for $W^+W^-$ final state. Note that the fluctuation of $\sigma_\text{off-shell}(m_a)$ lies within the Monte Carlo uncertainties and is likely due to numerical artifacts.} 
	\label{fig:xsec_prod_muon}
\end{figure}

To evaluate different ALP production mechanisms at a muon collider, their rates at $\sqrt{s}$=14~TeV are calculated by the full ME method without any parton-level cuts. For every $m_a$, we take $\{C_{WW}, C_{BB}, C_{BW}\}/f_a = \{1,1,0\}~\text{TeV}^{-1}$. From the results in Fig.~\ref{fig:xsec_prod_muon}, the VBF mechanism (corresponding to the blue curve) is the most efficient way to produce heavy ALPs at a muon collider, with the largest rate for any $m_a<\sqrt{s}$. Meanwhile, the associated $Va$ production rate remains small in the entire mass range considered. However, the SM tri-boson backgrounds for the $Va$ channel are also small. Therefore, the associated production mechanism could still be a relevant search channel. For the indirect search, the difference in the VBS rate induced by non-resonant ALP diagrams, namely $\sigma({\mu^+\mu^-}\to W^+W^-+X\text{'s})|_{ \text{off-shell}+{\rm SM}}-\sigma({\mu^+\mu^-}\to W^+W^-+X\text{'s})|_{\rm SM}$, is also plotted. Here $\sigma({\mu^+\mu^-}\to W^+W^-+X\text{'s})|_{\text{off-shell}+{\rm SM}}$ is the cross section of the same process including $t$-channel and off-shell $s$-channel ALP diagrams. Similar to the discussion in Section~\ref{ssec:indirect} for the LHC, the indirect search channels are not as useful as the direct ones due to the absence of narrow ALP resonances, though their rates could be larger than those of $Va$ channels. 

Although most of the beam energy is carried by the muon, a fraction of it could be transferred to collinear vector bosons and neutrinos. Such initial-state-radiation (ISR) effect introduces corrections to $\mu^+\mu^-$ annihilations. With ISR, the effective center-of-mass energy of $\mu^+\mu^-$ annihilation drops below $\sqrt{s}$, leading to a mild increase of event rates. Since the built-in muon PDF for muon beam is not yet available for our simulation framework, we approximate the ISR effect by inclusive collinear photon emissions. In particular, events with at most one extra photon $\gamma_\text{col}$ are simulated for both signals and backgrounds, \textit{i.e.}, $\mu^+\mu^-\to a\gamma+(\gamma_\text{col})$ and $\mu^+\mu^-\to 3\gamma+(\gamma_\text{col})$. As we are interested in cases where the ISR photon is unidentified, the extra ISR photon must satisfy $p_T < \SI{0.1}{\GeV}$ and $\abs{\eta} > 2.5$. For other photons, only a $p_T(\gamma)>10$~GeV cut is imposed at the parton level.

\subsection{Simulating Major SM Backgrounds}
\label{sec:smbkg}

\begin{table}
	\centering
	\begin{tabular}{l c c c}
		\hline\hline
		\multicolumn{1}{c}{Background Processes} & Parton-level Cuts & Methods & \multicolumn{1}{c}{$\sigma_\text{BKGD}\;(\si{\femto\barn})$}  \\
		\hline
		$\mu^+ \mu^- \to \gamma \gamma +X\text{'s}$ & \multirow{6}{*}{\parbox{3cm}{\centering 
				$p_{T,\gamma} \geq \SI{10}{\GeV}$ \\ 
				\& $p_{T,\ell} \geq \SI{0.5}{\GeV}$ \\ 
				\& $\abs{\eta_\ell} \leq 8$ 
		}} & \multirow{6}{*}{ME} & 1.05$\times 10^2$ \\
		\cline{1-1}\cline{4-4}
		$\mu^+ \mu^- \to ZZ(2\mu^+2\mu^-) +X\text{'s}$ & & & 0.310 \\
		\cline{1-1}\cline{4-4}	
		$\mu^+ \mu^- \to ZZ(4j) +X\text{'s}$ & & & 1.27$\times 10^2$ \\
		\cline{1-1}\cline{4-4}
		$\mu^+ \mu^- \to \gamma Z +X\text{'s}$ & & & 3.18$\times 10^2$  \\ 
		\cline{1-1}\cline{4-4}
		$\mu^+ \mu^- \to \gamma W^{\pm} +X\text{'s}$ & & & 15.8 \\
		\cline{1-1}\cline{4-4}
		$\mu^+ \mu^- \to W^{+} W^{-} +X\text{'s}$ & & & 6.26$\times 10^3$ \\
		\hline
		$\mu^+ \mu^- \to Z(\mu^+\mu^-) Z(jj) +X\text{'s}$ & \multirow{3}{*}{\parbox{3cm}{\centering no cut}} & ME + $m_\mu$ & 12.5 \\
		\cline{1-1}\cline{3-4}
		$\mu^\pm \gamma^* \to Z(\mu^+\mu^-) W^{\pm}(jj) +X\text{'s}$ & & \multirow{2}{*}{ME + EVA} & 26.6 \\
		\cline{1-1}\cline{4-4}
		$\mu^\pm \gamma^* \to Z(jj) W^{\pm}(jj) +X\text{'s}$ & & & 5.33$\times 10^2$ \\
		\hline\hline
	\end{tabular}
	\caption{Summary of cuts and cross sections for SM backgrounds simulated by \textsc{MadGraph}: $\gamma^*$ denotes a partonic photon from the muon beam, and $X$'s denote the beam remnants, including muon, muon neutrino, and their antiparticles. The background cross section $\sigma_{\text{BKGD}}$ is understood to include branching fractions to all secondary decays indicated in the parentheses. Decay products from $Z$'s and $W$'s are not subject to parton-level cuts. }
\label{tab:bkgXsec}
\end{table}

SM backgrounds are more involved to simulate compared to ALP signals. At a muon collider, the irreducible backgrounds for VBF ALP channels are the SM diboson processes, especially the VBS~\cite{Buarque:2021dji} due to large EW boson PDFs and sizable SM gauge couplings. The situation is analogous to the $gg\to jj$ backgrounds for the dijet resonance search at hadron colliders. The diboson invariant mass distributions of VBS backgrounds are continuous, on top of which we search for the sharp peak of an ALP resonance. The overall background level can thus be modeled by a background fit. Moreover, vector bosons from ALP decays give characteristic final states, \textit{e.g.}, a hard photon, a lepton pair at the $Z$ pole, or a fat jet with mass $\approx m_{W/Z}$, which further suppress reducible backgrounds like $\mu^+\mu^-\to q\bar{q}$.

In contrast to signal simulations, the soft and collinear singularities in VBS amplitudes require parton-level cuts, which render the Monte Carlo integrations convergent. However, stringent parton-level cuts on beam remnants could lead to background underestimations. It is necessary to choose proper cuts to achieve realistic background distributions and rates without oversampling in the divergent part of the phase space. Details of each background channel simulated are summarized in Table~\ref{tab:bkgXsec}.

Depending on the beam remnants, $X$'s, VBS backgrounds could be classified into three types: {\it 1)} $X$'s contain only muon neutrinos, with $W^+W^-$ as the corresponding initial-state vector bosons; {\it 2)} $X$'s contain only muons, with the initial states being $\gamma\gamma$, $Z\gamma$ or $ZZ$;  {\it 3)} $X$'s contain one muon and one muon neutrino, with the initial states $WZ$ or $W\gamma$. For type {\it 1)}, the background rates are not sensitive to cuts on neutrino beam remnants. For type {\it 2)}, the final state could only be $W^+W^-$ due to the SM gauge symmetry. Equivalently, $ZZ, Z\gamma, \gamma\gamma$ final states could only be from type {\it 1)}. The process initiated by $\gamma\gamma$ are susceptible to variations of parton-level cuts due to a collinear divergence. Its cross section is about one order of magnitude above the other processes in this category. Fortunately, at the cost of a long integration time, the ME method still provides plausible distributions of $\gamma\gamma$-initiated processes. In practice, we use the same parton-level cuts as in type {\it 1)}.

For type {\it 3)}, \textit{i.e.}, VBS $ZW$ and $\gamma W$ events, the final state dibosons have a net electric charge. They only contribute to the reducible backgrounds when the $W$ decays hadronically and is misidentified as a $Z$. For $\mu^+\mu^- \to \gamma W +X\text{'s}$ background, the ME method with soft parton-level cuts can give converging results. However, the ME method leads to numerical instability when generating $\mu^+\mu^- \to Z W +X\text{'s}$ samples. In this case, we use the effective vector boson approximation (EVA) based on the improved Weizsäcker-Williams formula~\cite{vonWeizsacker:1934nji,Williams:1934ad}, incorporated in \textsc{MadGraph}~\cite{Ruiz:2021tdt} to generate $\mu^\pm \gamma^* \to Z W +X\text{'s}$ samples, where $\gamma^*$ denotes the partonic photon from the muon beam. The vertex of $\mu\to W\nu$ emission is still handled by the ME calculation in this case. For simplicity, the $Z$ boson component in the muon beam is ignored, as its contribution to the overall VBS $ZW$ rate is negligible. For our background simulation, we set the factorization scale $Q = \sqrt{s}/2$. We also calculate the $ZW$ background rates with $Q = \sqrt{s}$ and $Q = \sqrt{s}/4$ and estimate the factorization scale induced uncertainty to be $\sim 6\%$.

The discussions above focus on the VBS backgrounds, the leading ones for ALP searches in the VBF channels. For the search in the associated production channel, we need to consider the SM tri-boson background. In particular, we will consider the $3\gamma$ final state, the simplest one in the associated production channel. Most VBS $3\gamma$ background can be removed by requiring the invariant mass of $3\gamma$ to be $\approx \sqrt{s}$. Simulation shows that the VBS $3\gamma$ background rate after applying the parton-level cut $m_{3\gamma}\in [0.9,1.0] \sqrt{s}$ is four orders of magnitude below that of $\mu^+\mu^-$ annihilations and can be ignored safely. The ISR effects are also included by the same method described in Section~\ref{ssec:signals}, giving a cross section of $\mu^+\mu^- \to 3 \gamma +(\gamma_{\rm col})=\SI{0.44}{\femto\barn}$.

\section{Muon Collider Phenomenology: VBF Channels \label{sec:VBF}}

\begin{table}[h!]
	\centering
	\begin{tabular}{c c c c c c}
		\hline\hline
		Final States & $p_T\;(\si{\GeV})$ & $m_{\ell\ell}$ & $R_{J}$ & $m_{J}\;(\si{\GeV})$ & Resonance Window \\
		\hline
		$\gamma\gamma$ & $\gamma_1: \geq 350$ & --- & --- & --- & $[m_a-3 \sigma_m, m_a+3 \sigma_m]$ \\
		\hline 
		\multirow{2}{*}{$\gamma Z(\ell \ell)$} & $\gamma_1 \geq 350$ & \multirow{2}{*}{$m_Z \pm 5 \Gamma_Z$}  & \multirow{2}{*}{---} & \multirow{2}{*}{---} & \multirow{2}{*}{$[m_a-3 \sigma_m, m_a+3 \sigma_m]$} \\
		& $\ell_1: \geq 100$ & & & & \\
		\hline\\[-12 pt]
		\multirow{2}{*}{$\gamma Z(jj)$} & $\gamma_1: \geq 350$ & \multirow{2}{*}{---} & \multirow{2}{*}{1.2} & \multirow{2}{*}{${m_Z}^{+ 2.5}_{- 1.5} \Gamma_Z$} & \multirow{2}{*}{$[m_a-1.5 \sigma_m, m_a+0.5 \sigma_m]$} \\
				& $J_1^\ast: \geq 200$ & & & \\
		\hline
		\multirow{2}{*}{$ZZ(4\ell)$} & $\ell_1: \geq 300$ & \multirow{2}{*}{$m_Z \pm 5 \Gamma_Z$} & \multirow{2}{*}{---} & \multirow{2}{*}{---} & \multirow{2}{*}{$[m_a-3 \sigma_m, m_a+3 \sigma_m]$} \\
		& $\ell_2: \geq 200$ & & & \\
		\hline
		\multirow{2}{*}{$ZZ(2\ell 2j)$} & $\ell_1: \geq 300$ & \multirow{2}{*}{$m_Z \pm 5 \Gamma_Z$} & \multirow{2}{*}{1.2} & \multirow{2}{*}{${m_Z}^{+ 2.5}_{- 1.5} \Gamma_Z$} & \multirow{2}{*}{$[m_a-1.5 \sigma_m, m_a+0.5 \sigma_m]$}\\
		& $J_1: \geq 200$ & & & \\
		\hline\\[-9.5pt]
		$ZZ(4j)$ & $J_{1,2}: \geq 300$ & --- & 1.0 & ${m_Z}^{+ 5}_{- 2} \Gamma_Z$ & $[m_a-1.5 \sigma_m, m_a+0.5 \sigma_m]$ \\\\[-9.5pt]
		\hline \\[-9.5pt]
		$WW(4j)$ & $J_{1,2}: \geq 300$ & --- & 1.0 & $ {m_W}_{- 5}^{+ 3} \Gamma_W$ & $[m_a-1.5 \sigma_m, m_a+0.5 \sigma_m]$ \\\\[-9.5pt]
		\hline\hline
	\end{tabular}
	\caption{Detector-level cuts imposed on various analyses: --- means that no value is applicable; $J$ represents a detector-level fat jet, while $j$ corresponds to a parton-level jet. The subscript of a particle name denotes its detector-level $p_T$ ordering. The jet radius $R=1$ is used for final states with two $J$'s (corresponding to 4$j$'s), and $R=1.2$ is used for final states with one $J$ ($2j$'s). The superscript $\ast$ indicates that the requirement is only for fat jets that do not contain photons. The width of each resonance peak is characterized by $\sigma_m$, see the main text for more details. }
	\label{tab:exc_cuts}
\end{table}

As discussed in Section~\ref{sec:HLLHC} and~\ref{sec:Simulation}, VBF production of ALPs provides a heavy diboson resonance with a narrow width. Therefore, we focus on the diboson final states which are fully visible, \textit{i.e.}, no $W\to \ell \nu$ or $Z\to \nu\bar{\nu}$ decays. All final states containing $\tau$'s will also be excluded from our analysis since they always contain neutrinos. Here we propose several analyses targeting all four ALP decay modes and their final states. Details of each analysis are described in Section~\ref{ssec:analysis}.

The general procedures of our analyses are straightforward. We first select events containing the same number of photons, leptons, and/or jets as the target ALP final state. Hard cuts on $p_T$ of $\mathcal{O}(m_a)$ are often applied to these objects at this stage to suppress SM VBS backgrounds. $Z/W$ reconstruction cuts are applied if the final state involves $Z/W$ decays. Sensitivities on $\sigma(\mu^+\mu^-\to a+X{\text{'s}})\times \text{BR}(a\to ZZ,~Z\gamma~,\gamma\gamma,~WW)$ are deduced from the signal and background yields within a narrow invariant mass window around $m_a$. The width of each invariant window is determined by the final state, the detector resolution, and the ALP mass. As to be discussed in Section~\ref{ssec:modelindep}, different initial states of VBF production only cause mild variations in signal efficiencies; hence, results provided in this section are approximately model independent. On the other hand, forward-region observables, \textit{e.g.}, number and energies of forward muons, are highly correlated with the initial boson states. Including forward-region information into the analysis may further enhance the sensitivity; see Section~\ref{ssec:forward}. Since the forward detector at a very-high-energy muon collider is still under development, we present limits utilizing forward-region information only as suggestive values.

\begin{figure}[h!]
\begin{center}
\includegraphics[width=0.8\linewidth]{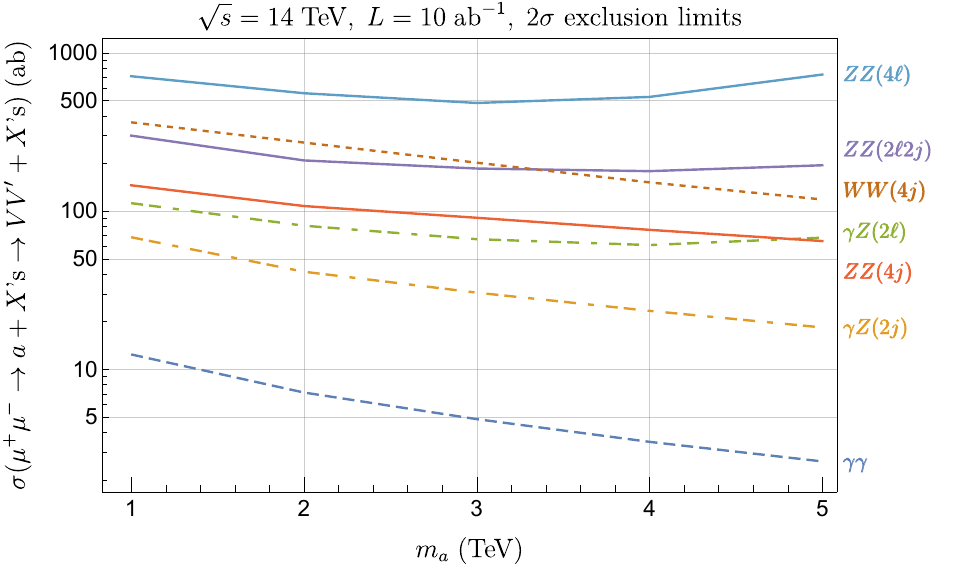}
\end{center}
\caption{Projected 2$\sigma$ exclusion limits on the VBF production cross section times the branching fraction of the ALP to a specific diboson final state, at a 14-TeV muon collider with $L=10$ ab$^{-1}$. Solid lines are exclusion limits from the analyses of $a \to ZZ$ with different secondary decays of $Z$'s, short dashed line is the exclusion limit from the analysis of $a\to WW$ with $W$'s decaying hadronically, long dashed line is the exclusion limit from the $\gamma\gamma$ channel, and exclusion limits from $a \to \gamma Z$ analyses are given by the dot-dashed lines. The $W$ and $Z$ decay products are indicated in the parentheses. }
\label{fig:IndependentLimit_14}
\end{figure}

\subsection{Event Reconstruction and Analyses}
\label{ssec:analysis}

\begin{figure} 
	\centering
	\includegraphics[width=\linewidth]{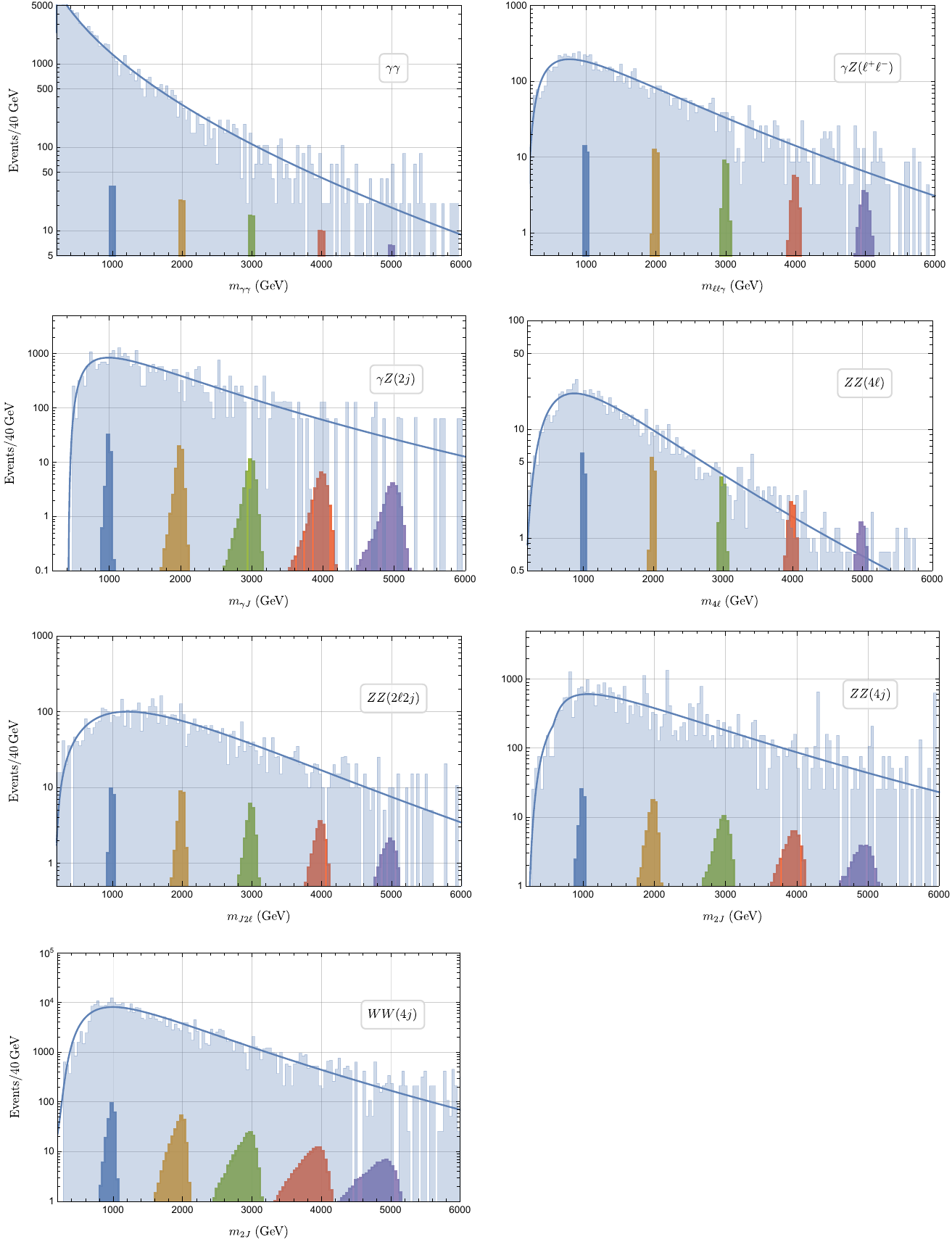}
	\caption{SM backgrounds~\textit{vs.}~ALP resonance signals for various ALP masses ($m_a$ = 1, 2, 3, 4, 5 TeV) at a $14$-TeV muon collider with $L = \SI{10}{\per\atto\barn}$. Simulated data are shown in histograms with a bin width of $\SI{40}{\GeV}$ while fitted background distributions are shown as solid curves. For all signal peaks, the cross sections are normalized to their $2\sigma$ exclusion limits.}
	\label{fig:MUON_ExcPlot}
	\end{figure}

To show concrete kinematic distributions and efficiencies, in the remainder of this section we will mainly present distributions and results for the $\sqrt{s}=$14~TeV benchmark. Analyses on seven final states are demonstrated in the following paragraphs, covering all four ALP diboson decay modes. Table~\ref{tab:exc_cuts} summarizes the key cut parameters for all the analyses. For convenience, we use $J$ to represent detector-level fat jets, while $j$ corresponds to parton-level jets. Also, the subscript of a particle name denotes its $p_T$ ordering at the detector level. The projected 2$\sigma$ exclusion limits on $\sigma(\mu^+\mu^-\to a +X{\text {'s}} \to VV^\prime+X{\text {'s}})$ are plotted in Fig.~\ref{fig:IndependentLimit_14}, assuming Poisson statistics~\cite{Cowan:2010js}. Since the typical signal and background yields are sufficiently larger than one, signal significance is well approximated by the Gaussian limit $S/\sqrt{S + B}$, in which $S$ ($B$) denotes the signal (background) counts in the signal region. Depending on ALP reconstruction efficiencies, EW boson decay branching ratios, and background levels, the limits from different analyses can vary by two orders of magnitude. Combining possible final states, the $a\to\gamma\gamma$ mode has the highest sensitivity, followed by $a\to Z\gamma$, $a\to ZZ$, and $a\to WW$ with the least sensitivity.

\paragraph{Analysis of $a\to \gamma\gamma$}
As the preselection criteria, only events containing exactly two isolated photons with the transverse momentum of the leading photon satisfying $p_T(\gamma_1)>350$~GeV are kept. Also, no more than one charged track other than forward muons with $p_T>10$ GeV is allowed. The signal efficiency for the $m_a=1(5)$~TeV benchmark is $\sim 57(74) \%$ after the preselection. The same cuts keep only $8.5\%$ of VBS diphoton backgrounds since most of them have low $p_T(\gamma)$ and are vetoed by the hard $p_T(\gamma_1)$ cut.

The diphoton invariant mass ($m_{\gamma\gamma}$) distributions for the VBS SM background and ALP benchmarks are shown in the upper left panel of Fig.~\ref{fig:MUON_ExcPlot}. The typical widths of the $a\to \gamma\gamma$ signal peaks are $\lesssim 20$~GeV due to the high resolution on the photons' momenta. The signal region for each ALP benchmark is then defined as $\abs{m_{\gamma\gamma} - m_a} < 3 \sigma_{m_{\gamma\gamma}}$. Here the characteristic width $\sigma_{m\gamma\gamma}$ is the half width at half maximum (HWHM) of each signal peak fitted by the Breit-Wigner distribution. Background yields in each signal region are calculated from a background fit using a generalized gamma distribution.\footnote{We have also fitted the backgrounds with other common distributions, \textit{e.g.}, exponential, log-normal, and Pareto. It was found that generalized gamma distribution provides a similar if not better background fit in general.} rather than the event counts to avoid large fluctuations.\footnote{To improve the quality of the background fit, any event with $m_{\gamma\gamma} < 200\, {\rm GeV}$ is dropped. Similar requirements are applied to all other VBF channel analyses.} Thanks to the high reconstruction efficiency and low backgrounds, the 2$\sigma$ exclusion limit on $\sigma(\mu^+\mu^-\to a +X{\text {'s}} \to \gamma\gamma +X{\text {'s}})$ reaches a few to 10 ab for TeV-scale axion at a $\sqrt{s}=14$~TeV muon collider with $L=10$~ab$^{-1}$.

\begin{figure}[h!]
\centering
\includegraphics[width=0.8\textwidth]{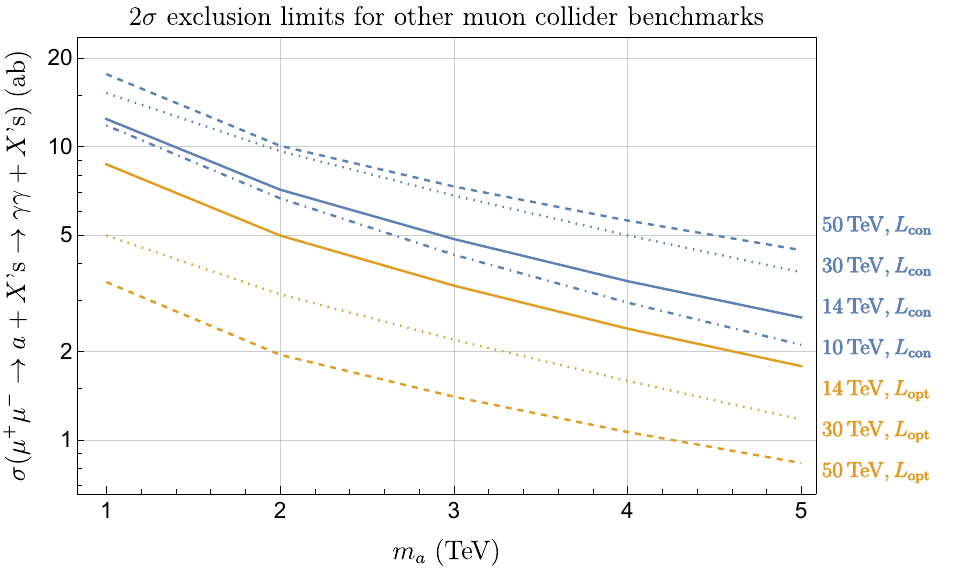}
\caption{2$\sigma$ limits on $\sigma(\mu^+\mu^-\to a +X{\text {'s}} \to \gamma\gamma +X{\text {'s}})$ at various muon collider benchmarks.}
\label{fig:diphoton_other_benchmarks}
\end{figure}

A similar search strategy applies to other benchmark beam energies. Fig.~\ref{fig:diphoton_other_benchmarks} shows the 2$\sigma$ limits on $\sigma(\mu^+\mu^-\to a +X{\text {'s}} \to \gamma\gamma +X{\text {'s}})$ for various muon collider energy benchmarks. Limits for optimistic luminosity scenarios are lower, corresponding to higher sensitivities, due to the aggressive increase of the integrated luminosities. However, in a conservative scenario where the luminosity is fixed, the exclusion limit becomes weaker for a higher $\sqrt{s}$. The reason is that as $m_a$ becomes much smaller than $\sqrt{s}$, the ALP produced picks up a greater boost in the beam direction on average. Consequently, the ALP decay products have larger $|\eta|$'s and are harder to reconstruct, reducing the signal efficiency. Meanwhile, SM backgrounds also increase moderately.

\paragraph{Analysis of $a\to \gamma Z(\to \ell\ell)$}

We first select events containing exactly one isolated photon with $p_T(\gamma)>350$~GeV, and two isolated leptons with the leading $p_T(\ell_{1})>100$~GeV. They must have opposite charges and same flavor. In addition, the dilepton invariant mass must satisfy $\abs{m_{\ell\ell} - m_{Z}} < 5 \Gamma_Z$ in which $\Gamma_Z$ denotes the full $Z$ width. All events containing more than one extra charged track with $p_T>10$ GeV are vetoed, excluding the lepton pairs discussed above and forward muons. 

The $m_{\ell\ell\gamma}$ distributions of signal and background samples are shown in the upper right panel of Fig.~\ref{fig:MUON_ExcPlot}. The signal peaks are also narrow, thanks to the high lepton and photon momentum resolutions. The signal region is defined as $\abs{m_{\ell\ell\gamma} - m_a} < 3 \sigma_{m\ell\ell\gamma}$, where $\sigma_{m\ell\ell\gamma}$ is the HWHM of the fitted signal peak. Although the signal reconstruction is straightforward and backgrounds are small, the $2\sigma$ upper limits on $\sigma (\mu^+ \mu^- \to a + X\text{'s}\to  \gamma Z + X\text{'s})$ from this analysis are only of $\mathcal{O}(100)$~ab because of the low $\text{BR}(Z \to \ell\ell) \approx 7\%$.

\paragraph{Analysis of $a\to \gamma Z(\to jj)$}

As the preselection, events with a hard isolated photon with $p_T(\gamma)>$ 350~GeV are chosen. Note that the inclusive Valencia jet algorithm may also recognize the isolated photon as a hard fat jet. Thus, the requirement on the number of fat jets with $R=1.2$ is relaxed to $N_J\in [1,2]$. For those events with two $J$'s, it is necessary to ensure that one $J$ comes from the hard photon. More specifically, we require one of them to have $m_J<\SI{1}{\GeV}$ and contain at most one charged track. In addition, the leading isolated photon must be inside this $J$. This light $J$ will be identified and treated as the photon instead of a hadronic $J$. The $p_T$ of the remaining $J$ (the true hadronic fat jet) must be greater than 200~GeV. To suppress $W\gamma$ backgrounds, its $m_J$ is further required to be $\in [m_Z - 1.5\Gamma_Z, m_Z + 2.5 \Gamma_Z]$. The $m_J$ window is asymmetric around $m_Z$ to suppress the $W\to jj$ background and provide adequate signal efficiencies. Numerically, the $m_J$ selection above suppresses the $W\gamma$ backgrounds by a factor of $\sim 3\%$ at the price of keeping $\sim 40\%$ of the signal events. Similar $W$ jet suppressions are also present in other analyses with hadronic $Z$ decays.

The middle left panel of Fig.~\ref{fig:MUON_ExcPlot} shows the $m_{\gamma J}$ distributions of SM backgrounds and several signal benchmarks. As expected, the finite jet momentum resolution makes signal peaks much wider than previous cases. Since signal peaks around $m_a$ are wide and asymmetric, the optimized signal region is defined as $m_{\gamma J}\in [m_a - 1.5 \sigma_{m\gamma J} , m_a+ 0.5 \sigma_{m\gamma J}]$ for each $m_a$. Here $\sigma_{m\gamma J}$ is simply the standard deviation of each signal peak, as the asymmetric peaks make the fitted HWHM less meaningful. Benefited from the large BR($Z\to jj$), the 2$\sigma$ limit on $\sigma (\mu^+ \mu^- \to a +X\text{'s} \to  Z\gamma +X\text{'s})$ from the $a\to \gamma Z(\to jj)$ analysis is at least a factor of two better than its $a\to \gamma Z(\to \ell\ell)$ counterpart, in the entire axion mass range we have studied. 

\paragraph{Analysis of $a\to ZZ(\to 4\ell)$}
For preselection, events with exactly four isolated leptons (excluding forward muons) with net electric charge zero are selected. Only those containing two same-flavor, opposite-sign lepton pairs both satisfying $\abs{m_{\ell^+\ell^-} - m_Z} \leq 5 \Gamma_Z$ are kept. To suppress the soft SM $ZZ\to 4\ell$ backgrounds, $p_T(\ell_{1,2})$ must be greater than 300 and 200~GeV, respectively. 

The resulting distributions of four lepton invariant mass, $m_{4\ell}$, are shown in the middle right panel of Fig.~\ref{fig:MUON_ExcPlot}. The narrow signal peaks leave a simple signal region definition as $\abs{m_{4\ell} - m_a} < 3 \sigma_{m_{4\ell}}$, where $\sigma_{m_{4\ell}}$ is the HWHM of each fitted signal peak. The projected $2\sigma$ upper bounds on $\sigma(\mu^+\mu^- \to a+ X{\text{'s}} \to ZZ+X{\text{'s}})$ are of $\mathcal{O}(1)$~fb, considerably weaker than other channels due to the small BR$(Z\to\ell\ell)^2\sim 5\times 10^{-3}$.

\paragraph{Analysis of $a\to Z(\to \ell\ell)Z(\to jj)$}

We first preselect events containing exactly two opposite-sign same-flavor leptons (excluding forward muons) with $p_T(\ell_1) > 300$~GeV. Their invariant mass must satisfy $\abs{m_{\ell^+\ell^-} - m_Z} \leq 5 \Gamma_Z$. A fat jet with $R=1.2$ and $m_J\in[m_Z-1.5 \Gamma_Z,m_Z+2.5 \Gamma_Z]$ is also required to suppress the reducible VBS $W(\to jj)Z(\to \ell\ell)$ background. The signal and background $m_{J2\ell}$ distributions are shown in the lower left panel in Fig.~\ref{fig:MUON_ExcPlot}. 

The signal regions are defined by $m_{J2\ell} \in [m_a - 1.5\sigma_{mJ2\ell}, m_a + 0.5\sigma_{mJ2\ell}]$, where $\sigma_{mJ2\ell}$ is the standard deviation of each asymmetric signal peak. Benefited from the high BR($Z\to jj$), the sensitivity provided by this analysis is stronger than the corresponding $a\to ZZ(4\ell)$ one.

\paragraph{Analysis of $a \to ZZ( \to 4j)$}

Events with exactly two fat jets ($R=1.0$) with both $p_T(J_{1,2})>300$~GeV are selected. Both $J$'s must satisfy $m_J\in[m_Z-2 \Gamma_Z,m_Z+5 \Gamma_Z]$, where a wider $m_J$ window is to ensure sufficient signal efficiencies. In addition, events containing more than one isolated leptons or photons with $p_T(\gamma/\ell)>10$~GeV are vetoed.   

The resulting distributions of two jet invariant mass are shown in the lower right panel of Fig.~\ref{fig:MUON_ExcPlot}. Finally, the signal region is defined by $m_{2J} \in [m_a - 1.5 \sigma_{m2J}, m_a + 0.5 \sigma_{m2J}]$ for each $m_a$ benchmark, where $\sigma_{m2J}$ is the standard deviation of the fitted signal peak. 
Benefited from the sizeable BR($Z\to jj$), the analysis provides the leading constraint on $ZZ$ detection channel with $\sigma(\mu^+\mu^- \to a+ X{\text{'s}} \to ZZ+X{\text{'s}})\lesssim \mathcal{O}(100)$~ab.

\paragraph{Analysis of $a\to WW (\to 4 j)$}
Similar to the previous analysis, events having two $R=1.0$ fat jets with both $p_T(J_{1,2})>300$~GeV are chosen. To veto VBS $ZZ(\to 4j)$ and $ZW(\to 4 j)$ backgrounds, each fat jet must satisfy $m_J \in [m_W - 5\Gamma_W, m_W+3\Gamma_W]$. Also, no more than one isolated lepton or photon with $p_T(\gamma/\ell) >10$ GeV is allowed. 

The $m_{JJ}$ distributions of signal and backgrounds of this analysis are shown in the last panel of Fig.~\ref{fig:MUON_ExcPlot}. The signal region is defined by $m_{2J}\in [m_a - 1.5 \sigma_{m2J}, m_a + 0.5 \sigma_{m2J}]$ for each $m_a$ benchmark, where $\sigma_{m2J}$ is the standard deviation of the signal peak. This analysis is subject to the highest background rate, stemming from the large VBS $WW$ cross section and BR($W\to jj)\sim 70\%$. Meanwhile, the reconstructed signal peaks are also the widest among all analyses due to the broad $m_J$ window. As a result, the constraint on $\sigma(\mu^+\mu^- \to a+ X{\text{'s}} \to WW+X{\text{'s}})$ is only about (100 - 400)~ab.

\subsection{Model Independence of the Limits}
\label{ssec:modelindep}
\begin{table}[h!]
\centering
\begin{tabular}{c >{$}c<{$\!\!} *{4}{>{$\!\!}c<{$\!\!}} >{\,\,\,}c<{\!\!\!} *{5}{>{\!\!\!}c<{\!\!\!}} >{\!\!\!}c}
\hline\hline
Model  &C_{WW}&:& C_{BB}&:& C_{BW} & BR$(\gamma\gamma)$ &:& BR$(\gamma Z)$ &:& BR$(ZZ)$ &:& BR$(WW)$ \\
\hline
Default & 1&:&1&:&0 & 7&:&6&:&22&:&65 \\
$\gamma\gamma$-phobic $WW$-phobic & 0&:&-1&:&1 & 0&:&63&:&37&:&0 \\
$\gamma\gamma$-phobic $WW$-philic 1 & -2&:&1&:&1 & 0&:&11&:&23&:&65 \\
$\gamma\gamma$-phobic $WW$-philic 2& -1&:&1&:&0 & 0&:&20&:&16&:&65 \\
$\gamma\gamma$-phobic $WW$-philic 3 & -1&:&0&:&1 & 0&:&5&:&31&:&64 \\
$\gamma\gamma$-philic $WW$-phobic & 0&:&1&:&1 & 89&:&0&:&10&:&0 \\
$\gamma\gamma$-philic  $\gamma Z$-phobic & 1&:&9.33&:&5.16 & \
87&:&0&:&0&:&13 \\
$\gamma\gamma$-philic $ZZ$-phobic & 1&:&3.33&:&4.33 & \
60&:&12&:&0&:&29 \\
Random model 1 & 1&:&1&:&1 & 15&:&12&:&11&:&61 \\
Random model 2 & 2.34&:&1&:&-2.26 & 0&:&3&:&33&:&64 \\
Random model 3 & 1&:&2.69&:&4 & 53&:&15&:&0&:&32 \\
Random model 4 & 1&:&1.54&:&-7.04 & 11&:&13&:&58&:&19 \\
\hline\hline
\end{tabular}
\caption{Details of ALP models in the model independence test. The branching fractions are normalized to percentages for readers' convenience.}
\label{tab:ModelDep_Sets}
\end{table}

The signal samples used to derive the limits summarized in Fig.~\ref{fig:IndependentLimit_14} are simulated assuming $C_{BB}= C_{WW}$ and $C_{BW} =0$. It is then necessary to check whether collider limits depend on a particular choice of ALP couplings. The concern is that with different combinations of couplings, the signal cut efficiency can be affected as the relative strengths in various production channels change. Even though multiple sets of EFT couplings allow ALPs to decay to the same final states, they can still affect the kinematics of produced ALPs as the importance of different initial states varies. Different initial states could lead to distinctive ALP momentum distributions, which may lead to different signal efficiencies in multiple analyses. For example, the ALPs produced in a $\gamma\gamma$-philic model ($i.e.$, the ALP coupling to $\gamma\gamma$ is significantly larger than the others after EWSB) will have a greater average longitudinal boost than the ALPs produced in a $WW$-philic model. Then, in the $\gamma\gamma$-philic model, the average $|\eta|$ of the ALP decay products will also be larger, causing more difficult event reconstructions and reduced signal efficiencies.

To examine the generality of limits in Section~\ref{ssec:analysis}, we survey twelve models parameterized by $\{C_{BB}, C_{WW}, C_{BW}\}$ and compare their signal efficiencies. The descriptions of tested models are provided in Table~\ref{tab:ModelDep_Sets}. The relative standard deviations of the signal efficiencies, \textit{i.e.} the standard deviations of signal efficiencies divided by the mean signal efficiencies of models considered, represent the analyses' model dependence\footnote{Note that all $VV'$-phobic models are dropped when evaluating the model dependence of a particular $VV'$ detection channel since there will be no such a decay.} and are collected in Table \ref{tab:ModelDep}. One could see that the signal efficiencies only vary by $\lesssim 20\%$ for all the analyses in a wide range of $m_a$. Therefore, it is reasonable to take the results in Fig.~\ref{fig:IndependentLimit_14} as model-independent limits for VBF produced pseudo-scalars at a 14-TeV muon collider. Although minor model-dependent variations remain, the exclusion limits in Fig.~\ref{fig:IndependentLimit_14} provide a solid order-of-magnitude estimate for further studies.

We also investigate some different new physics models with similar final state topologies. Specifically, we simulate models of heavy scalars with couplings to the square of EW field strengths~\cite{Ren:2014bza} and heavy vector boson with Chern-Simons couplings to EW gauge bosons~\cite{Antoniadis:2009ze,Bondarenko:2019tss}. It turns out that signal efficiencies of these scenarios match with the ALP models within the $1\sigma$ range shown in Table~\ref{tab:ModelDep}. Hence, it is very likely the upper limits in Fig.~\ref{fig:IndependentLimit_14} apply to generic VBF-produced resonances decaying to EW boson pairs.

\begin{table}[h!]
	\centering
	\begin{tabular}{c|*{7}{S[table-format=2.1]}}
		\hline\hline\\[-9.5pt]
		$m_a\;(\si{\TeV})$ & {$\gamma\gamma $} & {$\gamma Z(2\ell) $} & {$\gamma Z(2j) $} & {$ZZ(4\ell) $} & {$ZZ(2\ell 2j)$} & {$ZZ(4j)$} & {$WW(4j)$} \\
		\hline \\[-9.5pt]
		1 & 5. & 10.5 & 16.4 & 11. & 12.8 & 3.2 & 5.5 \\
		2 & 1.7 & 10.7 & 4.7 & 10.3 & 11.6 & 2.1 & 4.1 \\
		3 & 1.3 & 12.4 & 1.4 & 11.5 & 10.8 & 2.4 & 2.3 \\
		4 & 1. & 12.7 & 1.8 & 13.3 & 12.5 & 2.3 & 3.3 \\
		5 & 1.3 & 14.5 & 2.7 & 13.3 & 14.7 & 2.7 & 1.2 \\
		\hline\hline
	\end{tabular}
	\caption{Percentage variations of the signal efficiencies for different search channels, based on models listed in Table~\ref{tab:ModelDep_Sets}. The percentage variation is defined as the standard deviation of signal efficiencies divided by the mean signal efficiency of all the models we checked.}
	\label{tab:ModelDep}
\end{table}

\subsection{Analysis Including Forward Muons}
\label{ssec:forward}

One possible way to suppress the SM backgrounds $\mu^+\mu^- \to (ZZ, Z\gamma, \gamma\gamma) + X\text{'s}$ is to use the beam remnants' information in the forward region. As discussed in Section~\ref{sec:smbkg}, these processes are always associated with $W^+W^-$ initial state due to the SM gauge symmetry; therefore, $X$'s in these processes are only muon neutrinos. They will be strongly suppressed if we require one or two energetic forward muons in the analysis. Nevertheless, additional forward-region cuts also reduce signal efficiencies. Typically, only $\mathcal{O}(20\%)$ of signal events will remain if two forward muons are required. In addition, limits obtained with forward-region cuts are more sensitive to the model variation since different $C_{VV'}$ couplings in the EFT significantly affect the identity and kinematics of the beam remnants.

\begin{figure}
 	\begin{center}
 		\includegraphics[width=0.84\linewidth]{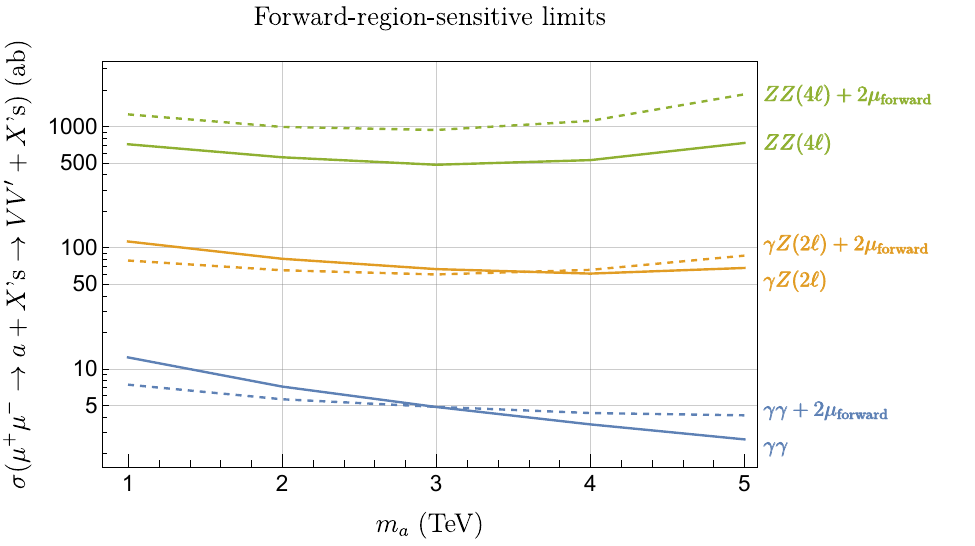}
 	\end{center}
 	\caption{Comparison between $2\sigma$ exclusion limits with (dashed lines) and without the forward-muon number cut (solid lines, the same as Fig.~\ref{fig:IndependentLimit_14}) for $\sqrt{s}=14$~TeV and $L = \SI{10}{\per\atto\barn}$. For cleaner reconstruction, we consider only channels with $\gamma$ or $Z\to \ell^+\ell^-$ final states.  }
 	\label{fig:ForwardLimit_14}
 \end{figure}

Exclusion limits using only events with two forward muons are shown in Fig.~\ref{fig:ForwardLimit_14}, together with corresponding ones without requiring forward muons from Fig.~\ref{fig:IndependentLimit_14}. In the high-$m_a$ region with small SM backgrounds already, results from analyses with no forward-muon cut take the lead as their signal efficiencies are higher. For some analyses, the forward-muon cuts could help improve the sensitivities slightly in the low-$m_a$ region. For example, requiring two forward muons improves the $a\to \gamma Z(\to 2\ell)$ and $a\to \gamma\gamma$ analyses when $m_a\lesssim 3$~TeV. However, due to the small SM background and decreased signal efficiency from the forward-muon cut, the limits from forward-muon-specific analyses are always weaker in the cleanest channel, \textit{i.e.}, $a\to ZZ(\to 4\ell)$.

\section{Muon Collider Phenomenology: Associated Production Channels \label{sec:Va}}

From Fig.~\ref{fig:xsec_prod_muon}, the associated production $\mu^+ \mu^- \to Va$ appears to be a subdominant production channel for ALP at a muon collider. However, this channel has several interesting features that could potentially benefit the detection and calls for further study. Firstly, as is clear from Fig.~\ref{fig:xsec_prod_muon}, $\sigma(\mu\mu\to Va)$ is approximately a constant for a wide range of $m_a$ up to the threshold $m_a\sim \sqrt{s}$, where VBF production is less efficient. Secondly, the major SM backgrounds for this channel are much smaller than the VBS backgrounds for the VBF channel~\cite{Han:2020pif,Costantini:2020stv,AlAli:2021let}.
Last but not least, the $Va$ production kinematics is almost independent of the ALP model choices when $\sqrt{s}\gg m_Z$. The limits derived are thus approximately model independent.

\subsection{Analysis of $a(\to \gamma\gamma)\gamma$}
\begin{figure}
	\centering
	\includegraphics[width=0.6\linewidth]{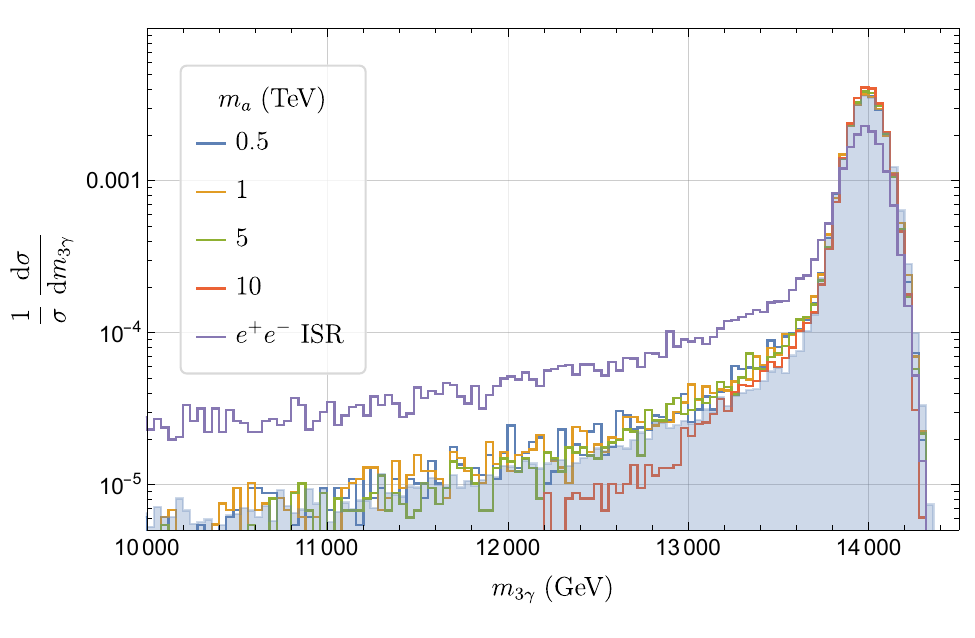}\\
	\includegraphics[width=\linewidth]{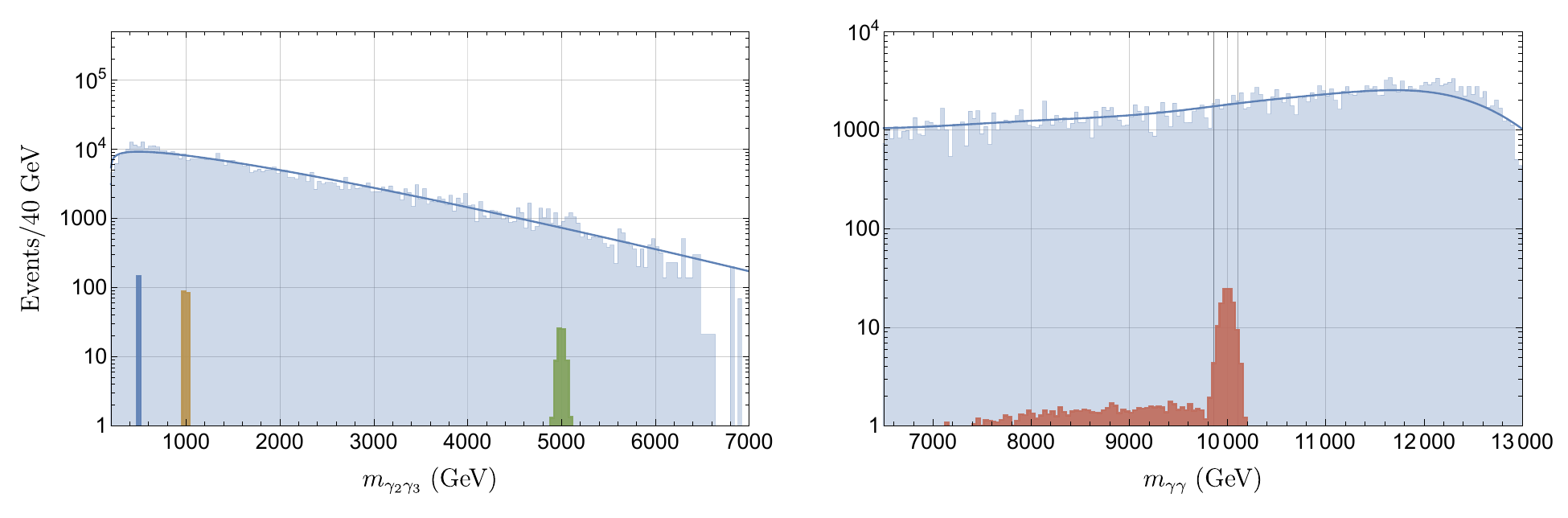}
	\caption{Distributions for $\gamma a(\gamma\gamma)$ associated production at a $14$-TeV muon collider with $L = \SI{10}{\per\atto\barn}$ data. In all panels, the light blue histograms represent the SM backgrounds. \textbf{Top}: Normalized distributions of reconstructed $m_{3\gamma}$. The smooth $m_{3\gamma}$ tail below $\sqrt{s}$ is the direct consequence of the undetected ISR photon. As a reference, $e^+ e^- \to 3 \gamma$ in SM with ISR at $\sqrt{s} = \SI{14}{\TeV}$ is also plotted. \textbf{Bottom left}: Reconstructed ALP peaks (for $m_a = 0.5, 1, 5$ TeV) and the SM background. Signals are scaled to the expected event counts when the production cross sections saturate the $2\sigma$ exclusion limits. \textbf{Bottom right}: reconstructed ALP peaks (for $m_a = 10$ TeV) over the SM background. The two vertical lines stand for the \textit{ad hoc} $m_{\gamma\gamma} \in [9.86, 10.1]\;\si{\TeV}$ window. }
	\label{fig:MUON_Va_ExcPlot}
\end{figure}

Inspired by the strong limit set by the VBF $a\to \gamma\gamma$ analysis, we focus on the $\mu^+ \mu^- \to \gamma a(\gamma\gamma)$ process as the benchmark of associated production channels. The top panel of Fig.~\ref{fig:MUON_Va_ExcPlot} displays the three photon invariant mass ($m_{3\gamma}$) distributions of the SM background and four representative $m_a$ benchmarks, simulated according to the method discussed in Section~\ref{sec:Simulation}. Note that $m_{3\gamma}$ can be regarded as a proxy for the energy scale of the hard process, sometimes denoted as $\sqrt{\hat{s}}$ in ISR-related studies. Aside from the detector smearing effects, the ISR-induced low-$m_{3\gamma}$ tails are also obvious. The $m_{3\gamma}$ distribution of the $e^+e^-\to 3\gamma$ process in SM with ISR effects (using the electron ISR PDF provided by \textsc{MadGraph5}) is also shown for comparison. Its ISR-induced tail is even more significant as expected since electrons radiate more than muons.

\begin{table}[h!]
	\centering
	\begin{tabular}{c c c c c}
		\hline\hline\\[-9.5pt]
		$m_a\;(\si{\TeV})$ & $0.5$ & $1$ & $5$ & $10$ \\
		\hline \\[-9.5pt]
		$\sigma_\text{Exc, no ISR}\;(\si{\atto\barn})$ & $28.6$ & $33.5$ & $44.4$ & $74.8$ \\
		\hline\\[-9.5pt]
		$\sigma_\text{Exc, with ISR}\;(\si{\atto\barn})$ & $31.6$ & $37.7$ & $48.6$ & $80.7$ \\
		\hline\hline
	\end{tabular}
	\caption{Exclusion limit at $2\sigma$ level for the tri-photon search at a 14-TeV muon collider with 10 ab$^{-1}$ data.}
		\label{tab:2sigmaVa}
\end{table}

Due to different kinematics, the analysis procedures for ``light ALP" ($m_a\leqslant\sqrt{s}/2$) and ``heavy ALP" ($m_a>\sqrt{s}/2$) benchmarks differ. For the light ALP case, we expect the associatively produced photon to be harder than the ALP decay products. We hence identify the two softer photons, \textit{i.e.}, $\gamma_2$ and $\gamma_3$, as ALP decay products. The detector-level cuts $p_{T}(\gamma_1) \geq \SI{4.5}{\TeV}$ and $\Delta R(\gamma_2\gamma_3) \leq 2$ are imposed to further suppress backgrounds with small $m_{\gamma_2\gamma_3}$. To suppress the potential VBF background, we also impose a detector-level cut on the hard scattering scale $m_{3\gamma} \geq 0.9 \sqrt{s}$. The lower left panel of Fig.~\ref{fig:MUON_Va_ExcPlot} shows the signal and background distributions of $m_{\gamma_2\gamma_3}$. Only events with a resonance peak and $m_{\gamma_2\gamma_3}\in [m_a- 3\sigma_{m\gamma_2\gamma_3},m_a+3\sigma_{m\gamma_2\gamma_3}]$ are selected, where $\sigma_{m\gamma_2\gamma_3}$ is the signal HWHM. The corresponding $2\sigma$ limits are shown in Table.~\ref{tab:2sigmaVa}. Notice that the continuous SM background level increases slowly when $m_{\gamma_2\gamma_3}$ decreases. Nevertheless, the larger background is compensated by the improved photon resolution (hence narrower $m_{\gamma_2\gamma_3}$ window) and higher signal efficiencies at low $m_{\gamma_2\gamma_3}$, rendering stronger constraints for smaller $m_a$. The limits obtained from samples without ISR are also shown for comparison. As expected, the ISR effect enhances the SM background rate and thus weakens the sensitivity slightly.

For the last benchmark ($m_a  > \sqrt{s}/2$), we use an angular-separation-ordered strategy instead. This is inspired by the fact that the associated photon is less energetic than the heavy ALP, and the heavy resonance has a small boost inside the detector. Therefore, the photon pair from the ALP decay shall have a larger angular separation, and the signal peak shall be reconstructed by the photon pair with the largest $\Delta R$. Our simulation also suggests that imposing $\min\{p_{T,\gamma_i}\} \geq \SI{1}{\TeV}$, $\max_{i,j}\{\Delta R_{\gamma_i \gamma_j}\} \leq 5$, and $m_{3\gamma} \geq 0.9 \sqrt{s}$ helps separate the signal and SM backgrounds. The $m_{\gamma\gamma}$ range of the signal region is determined \textit{ad hoc} as $m_{\gamma\gamma} \in [9.86, 10.1]\;\si{\TeV}$ since the reconstructed signal peak shape is highly non-trivial. Nonetheless, this choice of the signal region carves out most of the resonance peak around $m_a$. Relevant $m_{\gamma\gamma}$ distributions are shown in the lower right panel of Fig.~\ref{fig:MUON_Va_ExcPlot}, and the $2\sigma$ exclusion limit for the $m_a=10$~TeV benchmark is listed in Table~\ref{tab:2sigmaVa}.

To conclude, limits from the $a(\to \gamma\gamma)\gamma$ analysis are weaker than those of the VBF diphoton channel by a factor of ${\cal O}(1 -10)$ at a 14-TeV muon collider. A more detailed study on the ALP associated production channel with different final states and collider energies will be left to future work.

\section{Constraining the EFT} 
\label{sec:EFT}

With ALP production rates, branching ratios, and corresponding collider sensitivities known, one can put constraints on the ALP EFT in Eq.~\eqref{eq:EFT1}. ALP production rates, computed by \textsc{MadGraph5}, are quadratic functions of the EFT couplings, and the decay rates in terms of the couplings are given in Eq.~\eqref{eq:decay}. For collider sensitivities, we take the model-averaged limits discussed in Section~\ref{sec:VBF} and~\ref{sec:Va}.

\begin{figure}[h]
	\centering
	\includegraphics[width=0.65\linewidth]{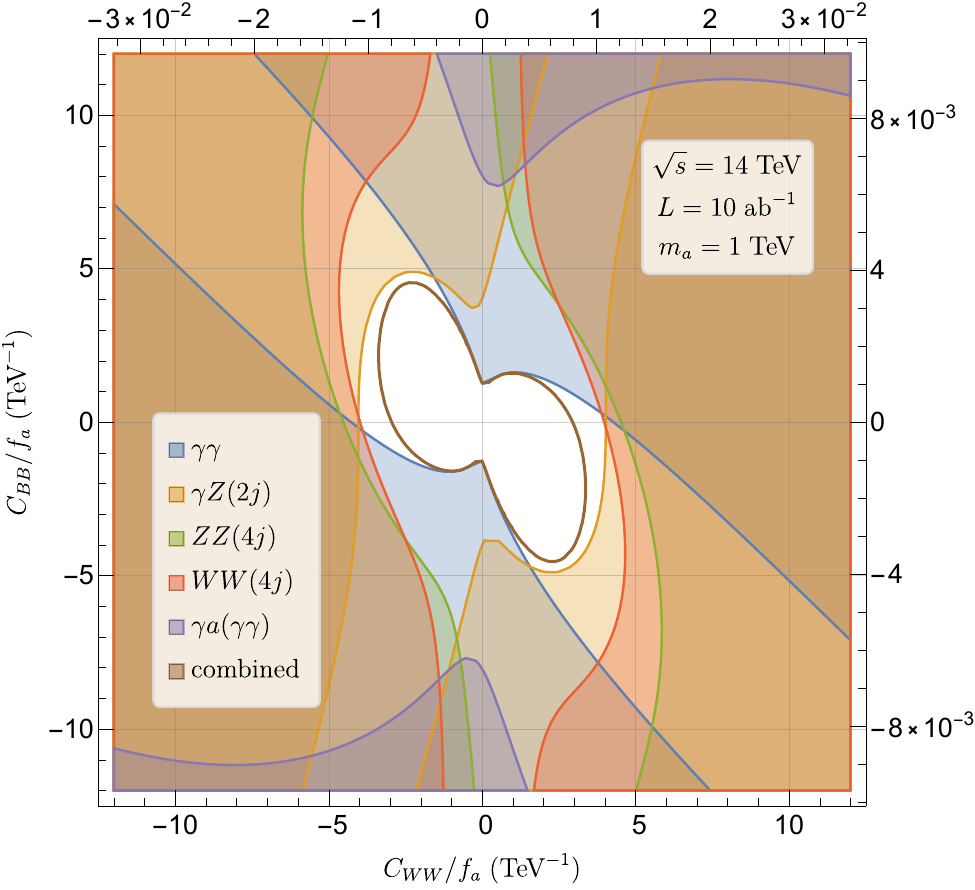}
	\caption{Projected $2\sigma$ constraint contours in the $C_{BB}/f_a-C_{WW}/f_a$ plane, for a $1$-TeV ALP at a $14$-TeV muon collider with $L =10$ ab$^{-1}$, assuming $C_{BW} = 0$. Detectable regions of different search channels are shaded with various colors. The solid brown butterfly-shaped curve around the center is the combined $2\sigma$ exclusion contour. Note that the bottom and left axes indicate values of the ALP couplings defined in Eq.~\eqref{eq:EFT1}, while the top and right axes indicate values of these couplings absorbing the gauge couplings and loop factors following the convention in Ref.~\cite{Brivio2017}.}
	\label{fig:MUON_no_CBW1}
\end{figure}

We first focus on the $C_{WW}$-$C_{BB}$ subspace with $C_{BW} = 0$, which is the most common scenario. The projected 2$\sigma$ constraints for a 1-TeV ALP at a 14-TeV muon collider with $L=10$ ab$^{-1}$ is shown in Fig.~\ref{fig:MUON_no_CBW1}. We show the parameter space covered by the four most sensitive VBF analyses and the $3\gamma$ analysis with associated production. While diphoton channel provides the strongest constraint on the signal rate, as demonstrated in Fig.~\ref{fig:MUON_ExcPlot}, it is insensitive to the $C_{BB} \simeq - C_{WW}$ direction in the coupling plane, as shown in Fig.~\ref{fig:MUON_no_CBW1}. This could be understood from Eq.~\eqref{eq:couplingsEWSB}: along this direction, the ALP coupling to photon vanishes. Thus, we need other analyses to complement the diphoton resonance search. In particular, the $\gamma Z(2j)$ analysis provides the strongest sensitivity along the $C_{BB} \simeq -C_{WW}$ direction. The combined $2\sigma$ contour takes a butterfly shape, as visible in Fig.~\ref{fig:MUON_no_CBW1}. In brief, a 14-TeV muon collider with $L=10$ ab$^{-1}$ data could probe $C_{BB}/f_a$ and $C_{WW}/f_a$ down to ${\cal O}$(1) TeV$^{-1}$, which is at least one order of magnitude more sensitive than HL-LHC!

\begin{figure}
	\centering
	\includegraphics[width=\linewidth]{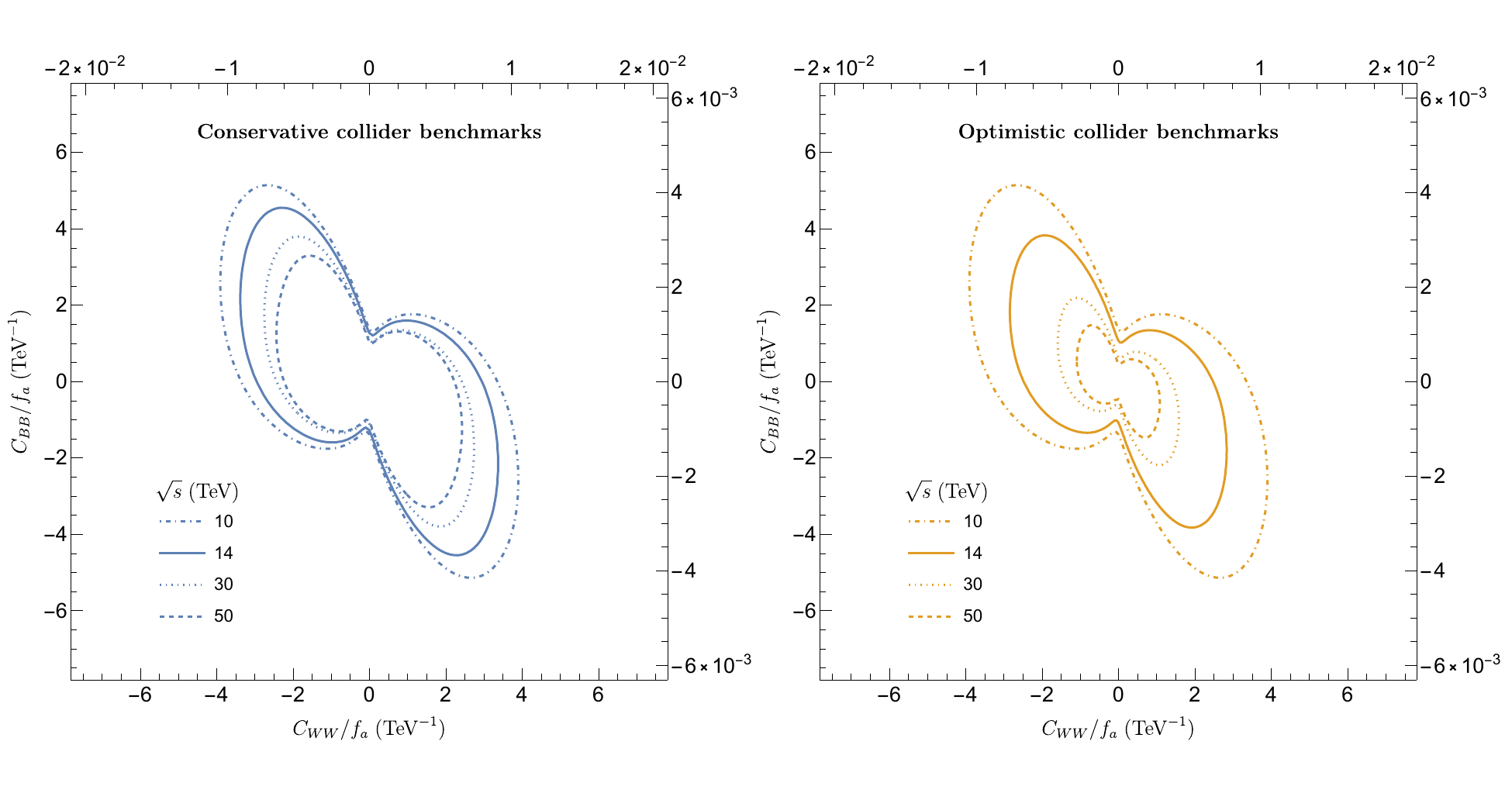}
	\caption{Projected $2\sigma$ constraint contours in the $C_{BB}/f_a-C_{WW}/f_a$ plane for different muon collider running benchmarks, assuming $C_{BW} = 0$. {\bf Left}: $L_{\rm con}$ scenario. {\bf Right}: $L_{\rm opt}$ scenario. }
	\label{fig:MUON_no_CBW2}
\end{figure}

Other muon collider energy/luminosity benchmarks are considered in Fig.~\ref{fig:MUON_no_CBW2}. The VBF $a\to\gamma\gamma$ exclusion limits for other running benchmarks are adopted from Section~\ref{ssec:analysis} directly. For limits on other final states, they are approximated by rescaling $\sigma(\mu^+\mu^-\to a +X{\text {'s}} \to \gamma\gamma +X{\text {'s}})$ limits in Fig.~\ref{fig:diphoton_other_benchmarks}. In particular, we assume that the ratios between the limits at any $(\sqrt{s}, L)$ and (14 TeV, 10 ab$^{-1}$) in every search channel stay the same as those from di-photon limits. We checked this assumption by completing the analysis with full simulations in the next-to-most-contributing detection channel, $\gamma Z(jj)$ channel, with a variety of values of $m_a$ and $\sqrt{s}$. The percentage variations between the rescaled exclusion limits and the exclusion limits with full simulations are $\lesssim 20\%$ for $\sqrt{s} =10,\,30$~TeV benchmarks and are at most $\sim 50\%$ for the $\sqrt{s}=50$~TeV benchmark. When $\sqrt{s}$ increases, the VBF ALP production rates increase, and consequently, the projected limits improve for both conservative and optimistic luminosity scenarios. The left panel of Fig.~\ref{fig:MUON_no_CBW2} shows the limits in the conservative luminosity scenario, while the right panel shows the ones in the optimistic scenario. By comparing both panels, the significant benefit from high integrated luminosities is evident. 


\begin{figure}
	\centering
	\includegraphics[width=0.5\textwidth]{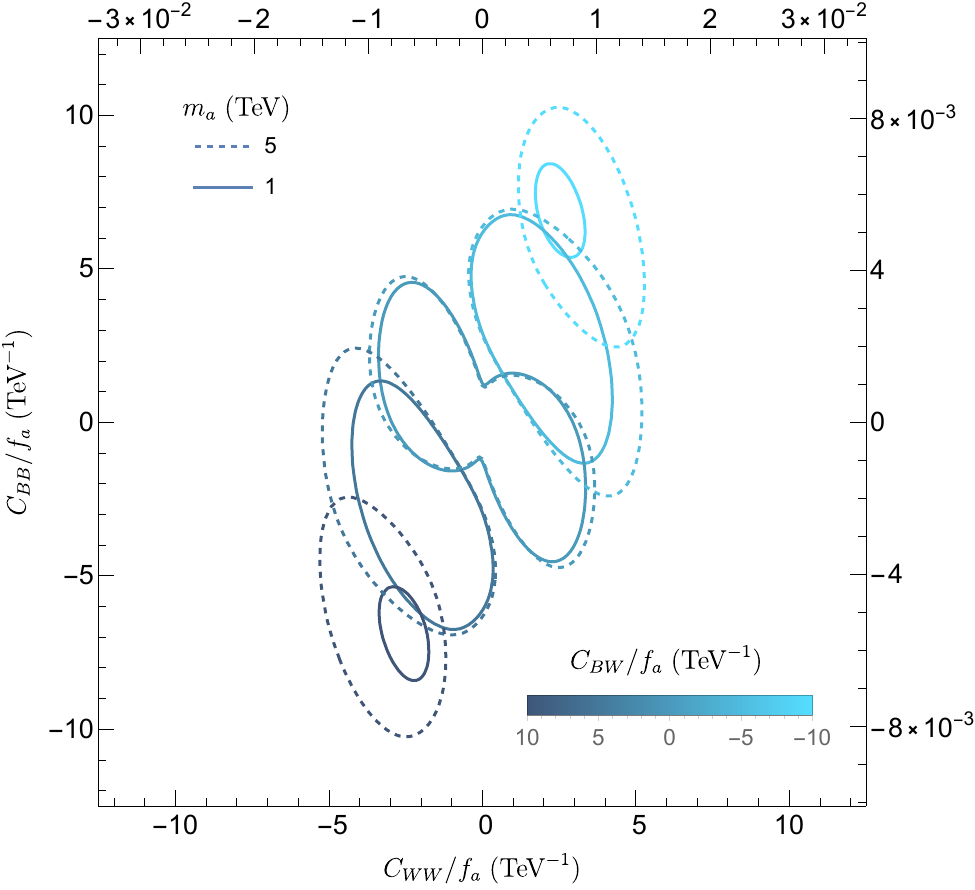} \\
	\includegraphics[width=0.39\textwidth]{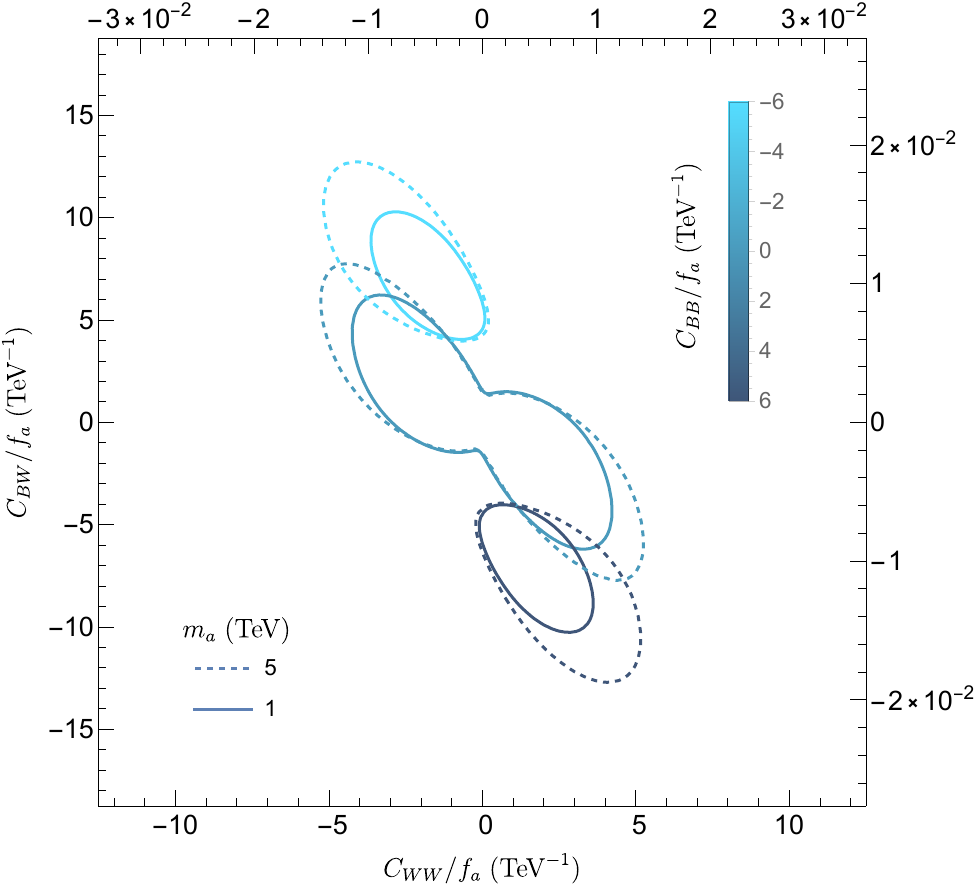}
	\includegraphics[width=0.39\textwidth]{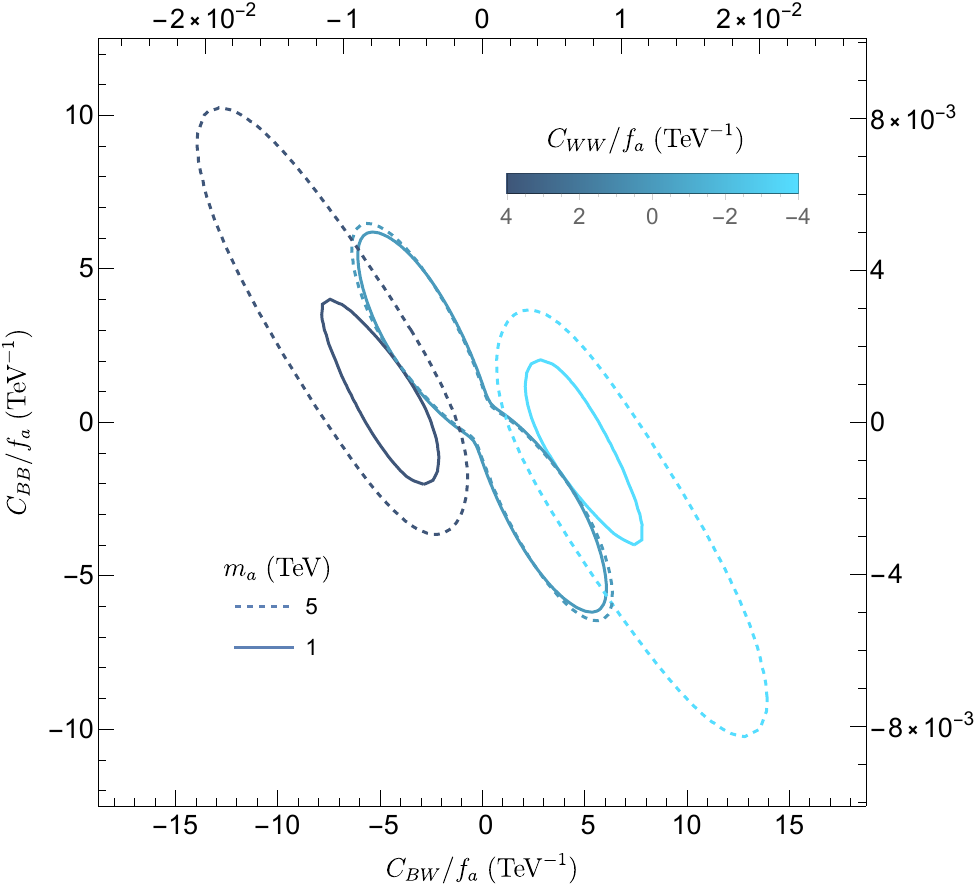}
	\caption{
		Combined projected $2\sigma$ constraints in various parameter subspaces at a 14-TeV muon collider with $L = 10$ ab$^{-1}$. Solid contours: $m_a = 1$ TeV. Dashed contours: $m_a = 5$ TeV. }
	\label{fig:MUON_with_CBW}
\end{figure}

Now we turn to the case in which $C_{BW} \neq 0$. We treat $C_{BW}$ on an equal footing as $C_{WW}$ and $C_{BB}$. To better understand the effects of a non-zero $C_{BW}$, we show exclusion contours in the $C_{WW}$-$C_{BB}$ plane with $C_{BW}/f_a = \{-10, -5, 0, 5, 10\} \; \si{\per\TeV}$, in the top panel of Fig.~\ref{fig:MUON_with_CBW}, with darker color indicating a higher value of $C_{BW}$. Near the origin of the plane, we obtain the familiar butterfly-shaped contour at the center of Fig.~\ref{fig:MUON_no_CBW1}. As $C_{BW}$ becomes non-zero, the constraints shift away from the origin due to the interference between operators. A larger $|C_{BW}|$ (so as the other couplings) also gives rise to a higher ALP production rate in general, leading to shrinking contours. 

Similar analyses are implemented in the $C_{WW}$-$C_{BW}$ plane with $C_{BB}/f_a = \{-6, 0, 6\} \; \si{\per\TeV}$, and in the $C_{BW}$-$C_{BB}$ plane with $C_{WW}/f_a = \{-4, 0, 4\}\; \si{\per\TeV}$. It is noteworthy that the overall constraints are the weakest in a direction $C_{WW}:C_{BB}:C_{BW}\simeq -3: -5: 8$ that simultaneously renders a small ALP production rate and elusive ALP decays with BR($a\to\gamma Z$)$\sim 5\%$ and BR($a\to\gamma\gamma)<1\%$.

\section{Summary and Outlook}
\label{sec:Conclusion}

In this article, we present a detailed analysis of TeV-scale ALP searches at a future high-energy muon collider. In particular, we focus on the searches that probe axion couplings to EW gauge bosons. The dominant  ALP production channel at a muon collider is the VBF channel due to the high virtual EW gauge boson content of high-energy muon beams, analogous to the virtual gluon content of high-energy proton beams. We show that the $a \to \gamma\gamma$ final state enjoys the highest sensitivity, followed by the $Z \gamma$ mode. We also analyze the associated production channel with a lower rate but a smaller background, which turns out to be less sensitive compared to the VBF diphoton channel. Meanwhile, we show the projected constraints on various subspaces of ALP couplings in the EFT. A muon collider with energy $\gtrsim 10$ TeV could improve the sensitivity to ALP couplings by at least one order of magnitude compared to the HL-LHC, as well as expand the mass range of ALPs that could be searched for. We also demonstrate that the model-independent limits on ALP VBF production and decays are also applicable to generic BSM resonances coupling to EW gauge bosons, which could benefit future related studies. 
This study serves as another example of the great physics potentials of a high-energy muon collider.

There are several directions to expand the work. First, we base our study on the EFT of an ALP coupling to EW gauge bosons. While the EFT usually serves as a useful model-independent theoretical framework for experimental searches, the UV completions could predict (model-dependent) degrees of freedom and signals that could also be within reach. In particular, the couplings of heavy ALPs that current and future colliders are sensitive to are pretty large. They could be induced by heavy fermions carrying EW charges, which could be searched for as well. It will be useful to survey and classify possible UV completions of the ALP EFTs, check whether there are some generic predictions for collider phenomenologies, and compare the sensitivities to the associated UV degrees of freedom and to ALPs. Secondly, the beam remnants in the forward region in the VBF channels ecode additional information. We carry out a crude estimate in this paper and find that they could lead to minor improvements of sensitivities in some mass range of ALPs. It will be useful to examine how to employ the forward-region information in a more sophisticated way, which could be valuable inputs for the design of a future muon collider. Thirdly, despite of the weaker constraints on the EFT couplings from $\gamma a(\gamma\gamma)$ associated production channel, a generic $Va$ channel is robust against model variations and systematic uncertainties. In this study, we use the most straightforward observables and analysis strategy on the benchmark tri-photon channel. Dedicated designs of observables and analyses in other $Va$ channels can potentially complement the VBF search at a future muon collider and are worth further investigations. 

\textit{Note added --}
After our preprint draft appeared on arxiv, we became aware of another work~\cite{Han:2022mzp} on the same subject. While both papers share some components in analyses, they differ and complement each other in several aspects.

\section*{Acknowledgements}

We thank Keping Xie for useful discussions. JF and LL are supported by the DOE grant DE-SC-0010010 and the NASA grant 80NSSC18K1010.

\bibliography{Ref}

\providecommand{\href}[2]{#2}\begingroup\raggedright\begin{thebibliography}{100}

\bibitem{Peccei:1977ur}
R.~D. Peccei and H.~R. Quinn, \emph{{Constraints Imposed by CP Conservation in
  the Presence of Instantons}},
  \href{https://doi.org/10.1103/PhysRevD.16.1791}{\emph{Phys. Rev. D}
  {\bfseries 16} (1977) 1791}.

\bibitem{Peccei:1977hh}
R.~D. Peccei and H.~R. Quinn, \emph{{CP Conservation in the Presence of
  Instantons}}, \href{https://doi.org/10.1103/PhysRevLett.38.1440}{\emph{Phys.
  Rev. Lett.} {\bfseries 38} (1977) 1440}.

\bibitem{Weinberg:1977ma}
S.~Weinberg, \emph{{A New Light Boson?}},
  \href{https://doi.org/10.1103/PhysRevLett.40.223}{\emph{Phys. Rev. Lett.}
  {\bfseries 40} (1978) 223}.

\bibitem{Wilczek:1977pj}
F.~Wilczek, \emph{{Problem of Strong $P$ and $T$ Invariance in the Presence of
  Instantons}}, \href{https://doi.org/10.1103/PhysRevLett.40.279}{\emph{Phys.
  Rev. Lett.} {\bfseries 40} (1978) 279}.

\bibitem{Kim:1979if}
J.~E. Kim, \emph{{Weak Interaction Singlet and Strong CP Invariance}},
  \href{https://doi.org/10.1103/PhysRevLett.43.103}{\emph{Phys. Rev. Lett.}
  {\bfseries 43} (1979) 103}.

\bibitem{Shifman:1979if}
M.~A. Shifman, A.~I. Vainshtein and V.~I. Zakharov, \emph{{Can Confinement
  Ensure Natural CP Invariance of Strong Interactions?}},
  \href{https://doi.org/10.1016/0550-3213(80)90209-6}{\emph{Nucl. Phys. B}
  {\bfseries 166} (1980) 493}.

\bibitem{Zhitnitsky:1980tq}
A.~R. Zhitnitsky, \emph{{On Possible Suppression of the Axion Hadron
  Interactions. (In Russian)}}, {\emph{Sov. J. Nucl. Phys.} {\bfseries 31}
  (1980) 260}.

\bibitem{Dine:1981rt}
M.~Dine, W.~Fischler and M.~Srednicki, \emph{{A Simple Solution to the Strong
  CP Problem with a Harmless Axion}},
  \href{https://doi.org/10.1016/0370-2693(81)90590-6}{\emph{Phys. Lett. B}
  {\bfseries 104} (1981) 199}.

\bibitem{Preskill:1982cy}
J.~Preskill, M.~B. Wise and F.~Wilczek, \emph{{Cosmology of the Invisible
  Axion}}, \href{https://doi.org/10.1016/0370-2693(83)90637-8}{\emph{Phys.
  Lett. B} {\bfseries 120} (1983) 127}.

\bibitem{Dine:1982ah}
M.~Dine and W.~Fischler, \emph{{The Not So Harmless Axion}},
  \href{https://doi.org/10.1016/0370-2693(83)90639-1}{\emph{Phys. Lett. B}
  {\bfseries 120} (1983) 137}.

\bibitem{Abbott:1982af}
L.~F. Abbott and P.~Sikivie, \emph{{A Cosmological Bound on the Invisible
  Axion}}, \href{https://doi.org/10.1016/0370-2693(83)90638-X}{\emph{Phys.
  Lett. B} {\bfseries 120} (1983) 133}.

\bibitem{Gaillard:2018xgk}
M.~K. Gaillard, M.~B. Gavela, R.~Houtz, P.~Quilez and R.~Del~Rey, \emph{{Color
  unified dynamical axion}},
  \href{https://doi.org/10.1140/epjc/s10052-018-6396-6}{\emph{Eur. Phys. J. C}
  {\bfseries 78} (2018) 972}
  [\href{https://arxiv.org/abs/1805.06465}{{\ttfamily 1805.06465}}].

\bibitem{Jaeckel:2012yz}
J.~Jaeckel, M.~Jankowiak and M.~Spannowsky, \emph{{LHC probes the hidden
  sector}}, \href{https://doi.org/10.1016/j.dark.2013.06.001}{\emph{Phys. Dark
  Univ.} {\bfseries 2} (2013) 111}
  [\href{https://arxiv.org/abs/1212.3620}{{\ttfamily 1212.3620}}].

\bibitem{Mimasu:2014nea}
K.~Mimasu and V.~Sanz, \emph{{ALPs at Colliders}},
  \href{https://doi.org/10.1007/JHEP06(2015)173}{\emph{JHEP} {\bfseries 2015}
  (2015) 173} [\href{https://arxiv.org/abs/1409.4792}{{\ttfamily 1409.4792}}].

\bibitem{Jaeckel:2015jla}
J.~Jaeckel and M.~Spannowsky, \emph{{Probing MeV to 90 GeV axion-like particles
  with LEP and LHC}},
  \href{https://doi.org/10.1016/j.physletb.2015.12.037}{\emph{Phys. Lett. B}
  {\bfseries 753} (2016) 482}
  [\href{https://arxiv.org/abs/1509.00476}{{\ttfamily 1509.00476}}].

\bibitem{Knapen:2016moh}
S.~Knapen, T.~Lin, H.~K. Lou and T.~Melia, \emph{{Searching for Axionlike
  Particles with Ultraperipheral Heavy-Ion Collisions}},
  \href{https://doi.org/10.1103/PhysRevLett.118.171801}{\emph{Phys. Rev. Lett.}
  {\bfseries 118} (2017) 171801}
  [\href{https://arxiv.org/abs/1607.06083}{{\ttfamily 1607.06083}}].

\bibitem{Bauer:2017ris}
M.~Bauer, M.~Neubert and A.~Thamm, \emph{{Collider Probes of Axion-Like
  Particles}}, \href{https://doi.org/10.1007/JHEP12(2017)044}{\emph{JHEP}
  {\bfseries 2017} (2017) 044}
  [\href{https://arxiv.org/abs/1708.00443}{{\ttfamily 1708.00443}}].

\bibitem{Craig:2018kne}
N.~Craig, A.~Hook and S.~Kasko, \emph{{The Photophobic ALP}},
  \href{https://doi.org/10.1007/JHEP09(2018)028}{\emph{JHEP} {\bfseries 2018}
  (2018) 028} [\href{https://arxiv.org/abs/1805.06538}{{\ttfamily
  1805.06538}}].

\bibitem{Lee:2018pag}
L.~Lee, C.~Ohm, A.~Soffer and T.-T. Yu, \emph{{Collider Searches for Long-Lived
  Particles Beyond the Standard Model}},
  \href{https://doi.org/10.1016/j.ppnp.2019.02.006}{\emph{Prog. Part. Nucl.
  Phys.} {\bfseries 106} (2019) 210}
  [\href{https://arxiv.org/abs/1810.12602}{{\ttfamily 1810.12602}}].

\bibitem{Bauer:2018uxu}
M.~Bauer, M.~Heiles, M.~Neubert and A.~Thamm, \emph{{Axion-Like Particles at
  Future Colliders}},
  \href{https://doi.org/10.1140/epjc/s10052-019-6587-9}{\emph{Eur. Phys. J. C}
  {\bfseries 79} (2019) 74} [\href{https://arxiv.org/abs/1808.10323}{{\ttfamily
  1808.10323}}].

\bibitem{Hook:2019qoh}
A.~Hook, S.~Kumar, Z.~Liu and R.~Sundrum, \emph{{High Quality QCD Axion and the
  LHC}}, \href{https://doi.org/10.1103/PhysRevLett.124.221801}{\emph{Phys. Rev.
  Lett.} {\bfseries 124} (2020) 221801}
  [\href{https://arxiv.org/abs/1911.12364}{{\ttfamily 1911.12364}}].

\bibitem{Carmona:2021seb}
A.~Carmona, C.~Scherb and P.~Schwaller, \emph{{Charming ALPs}},
  \href{https://doi.org/10.1007/JHEP08(2021)121}{\emph{JHEP} {\bfseries 2021}
  (2021) 121} [\href{https://arxiv.org/abs/2101.07803}{{\ttfamily
  2101.07803}}].

\bibitem{Wang:2021uyb}
D.~Wang, L.~Wu, J.~M. Yang and M.~Zhang, \emph{{Photon-jet events as a probe of
  axionlike particles at the LHC}},
  \href{https://doi.org/10.1103/PhysRevD.104.095016}{\emph{Phys. Rev. D}
  {\bfseries 104} (2021) 095016}
  [\href{https://arxiv.org/abs/2102.01532}{{\ttfamily 2102.01532}}].

\bibitem{Agrawal:2021dbo}
P.~Agrawal et~al., \emph{{Feebly-interacting particles: FIPs 2020 workshop
  report}}, \href{https://doi.org/10.1140/epjc/s10052-021-09703-7}{\emph{Eur.
  Phys. J. C} {\bfseries 81} (2021) 1015}
  [\href{https://arxiv.org/abs/2102.12143}{{\ttfamily 2102.12143}}].

\bibitem{Florez:2021zoo}
A.~Fl\'orez, A.~Gurrola, W.~Johns, P.~Sheldon, E.~Sheridan, K.~Sinha et~al.,
  \emph{{Probing axionlike particles with $\gamma\gamma$ final states from
  vector boson fusion processes at the LHC}},
  \href{https://doi.org/10.1103/PhysRevD.103.095001}{\emph{Phys. Rev. D}
  {\bfseries 103} (2021) 095001}
  [\href{https://arxiv.org/abs/2101.11119}{{\ttfamily 2101.11119}}].

\bibitem{Liu:2021lan}
Y.~Liu and B.~Yan, \emph{{Searching for the axion-like particle at the EIC}},
  \href{https://arxiv.org/abs/2112.02477}{{\ttfamily 2112.02477}}.

\bibitem{FCC:2018evy}
{\scshape FCC} collaboration, \emph{{FCC-ee: The Lepton Collider}: {Future
  Circular Collider Conceptual Design Report Volume 2}},
  \href{https://doi.org/10.1140/epjst/e2019-900045-4}{\emph{Eur. Phys. J. ST}
  {\bfseries 228} (2019) 261}.

\bibitem{CEPCStudyGroup:2018ghi}
{\scshape CEPC Study Group} collaboration, \emph{{CEPC Conceptual Design
  Report: Volume 2 - Physics \& Detector}},
  \href{https://arxiv.org/abs/1811.10545}{{\ttfamily 1811.10545}}.

\bibitem{Bambade:2019fyw}
P.~Bambade et~al., \emph{{The International Linear Collider: A Global
  Project}},  \href{https://arxiv.org/abs/1903.01629}{{\ttfamily 1903.01629}}.

\bibitem{CLICdp:2018cto}
{\scshape CLICdp, CLIC} collaboration, \emph{{The Compact Linear Collider
  (CLIC) - 2018 Summary Report}},
  \href{https://arxiv.org/abs/1812.06018}{{\ttfamily 1812.06018}}.

\bibitem{FCC:2018vvp}
{\scshape FCC} collaboration, \emph{{FCC-hh: The Hadron Collider}: {Future
  Circular Collider Conceptual Design Report Volume 3}},
  \href{https://doi.org/10.1140/epjst/e2019-900087-0}{\emph{Eur. Phys. J. ST}
  {\bfseries 228} (2019) 755}.

\bibitem{Skrinsky:1981ht}
A.~N. Skrinsky and V.~V. Parkhomchuk, \emph{{Cooling Methods for Beams of
  Charged Particles. (In Russian)}}, {\emph{Sov. J. Part. Nucl.} {\bfseries 12}
  (1981) 223}.

\bibitem{Neuffer:1983xya}
D.~Neuffer, \emph{{Principles and Applications of Muon Cooling}}, {\emph{Conf.
  Proc. C} {\bfseries 830811} (1983) 481}.

\bibitem{Neuffer:1986dg}
D.~Neuffer, \emph{{Multi-TeV muon colliders}},  in \emph{{AIP} Conf. Proc.},
  F.~E. Mills, ed., vol.~156, pp.~201--208, {AIP}, 1987,
  \href{https://doi.org/10.1063/1.36456}{DOI}.

\bibitem{Barger:1995hr}
V.~D. Barger, M.~S. Berger, J.~F. Gunion and T.~Han, \emph{{$s$-Channel Higgs
  Boson Production at a Muon-Muon Collider}},
  \href{https://doi.org/10.1103/PhysRevLett.75.1462}{\emph{Phys. Rev. Lett.}
  {\bfseries 75} (1995) 1462}
  [\href{https://arxiv.org/abs/hep-ph/9504330}{{\ttfamily hep-ph/9504330}}].

\bibitem{Barger:1996jm}
V.~D. Barger, M.~S. Berger, J.~F. Gunion and T.~Han, \emph{{Higgs Boson physics
  in the $s$ channel at $\mu^+ \mu^-$ colliders}},
  \href{https://doi.org/10.1016/S0370-1573(96)00041-5}{\emph{Phys. Rept.}
  {\bfseries 286} (1997) 1}
  [\href{https://arxiv.org/abs/hep-ph/9602415}{{\ttfamily hep-ph/9602415}}].

\bibitem{Ankenbrandt:1999cta}
C.~M. Ankenbrandt et~al., \emph{{Status of muon collider research and
  development and future plans}},
  \href{https://doi.org/10.1103/PhysRevSTAB.2.081001}{\emph{Phys. Rev. ST
  Accel. Beams} {\bfseries 2} (1999) 081001}
  [\href{https://arxiv.org/abs/physics/9901022}{{\ttfamily physics/9901022}}].

\bibitem{Snowmass}
Snowmass, \emph{Discussion forum on muon colliders},  2021.

\bibitem{Aime:2022flm}
C.~Aim\`e et~al., \emph{{Muon Collider Physics Summary}},  in \emph{{2022
  Snowmass Summer Study}}, 3, 2022,
  \href{https://arxiv.org/abs/2203.07256}{{\ttfamily 2203.07256}}.

\bibitem{Han:2020uid}
T.~Han, Y.~Ma and K.~Xie, \emph{{High Energy Leptonic Collisions and
  Electroweak Parton Distribution Functions}},
  \href{https://doi.org/10.1103/PhysRevD.103.L031301}{\emph{Phys. Rev. D}
  {\bfseries 103} (2021) l031301}
  [\href{https://arxiv.org/abs/2007.14300}{{\ttfamily 2007.14300}}].

\bibitem{Bertone:2017bme}
{\scshape NNPDF} collaboration, \emph{{Illuminating the photon content of the
  proton within a global PDF analysis}},
  \href{https://doi.org/10.21468/SciPostPhys.5.1.008}{\emph{SciPost Phys.}
  {\bfseries 5} (2018) 008} [\href{https://arxiv.org/abs/1712.07053}{{\ttfamily
  1712.07053}}].

\bibitem{Palmer:2014nza}
R.~B. Palmer, \emph{{Muon Colliders}},
  \href{https://doi.org/10.1142/S1793626814300072}{\emph{Rev. Accel. Sci.
  Tech.} {\bfseries 07} (2014) 137}.

\bibitem{Delahaye:2019omf}
J.~P. Delahaye, M.~Diemoz, K.~Long, B.~Mansouli\'e, N.~Pastrone, L.~Rivkin
  et~al., \emph{{Muon Colliders}},
  \href{https://arxiv.org/abs/1901.06150}{{\ttfamily 1901.06150}}.

\bibitem{Delahaye:2013jla}
J.-P. Delahaye et~al., \emph{{Enabling Intensity and Energy Frontier Science
  with a Muon Accelerator Facility in the U.S.: A White Paper Submitted to the
  2013 U.S. Community Summer Study of the Division of Particles and Fields of
  the American Physical Society}},  in \emph{{Community Summer Study 2013}:
  {Snowmass on the Mississippi}}, Aug., 2013,
  \href{https://arxiv.org/abs/1308.0494}{{\ttfamily 1308.0494}}.

\bibitem{Delahaye:2014vvd}
J.-P. Delahaye et~al., \emph{{A Staged Muon Accelerator Facility For Neutrino
  and Collider Physics}},  in \emph{{Proc. 5th IPAC}}, pp.~1872--1876, JACoW,
  July, 2014, \href{https://arxiv.org/abs/1502.01647}{{\ttfamily 1502.01647}},
  \href{https://doi.org/10.18429/JACoW-IPAC2014-WEZA02}{DOI}.

\bibitem{Ryne:IPAC2015-WEPWA057}
R.~Ryne et~al., \emph{{D}esign {C}oncepts for {M}uon{-B}ased {A}ccelerators},
  in \emph{Proc. 6th International Particle Accelerator Conference (IPAC'15),
  Richmond, VA, USA, May 3-8, 2015}, no.~6 in International Particle
  Accelerator Conference, (Geneva, Switzerland), pp.~2633--2636, JACoW, June,
  2015,
  \href{https://doi.org/https://doi.org/10.18429/JACoW-IPAC2015-WEPWA057}{DOI}.

\bibitem{Long:2020wfp}
K.~Long, D.~Lucchesi, M.~Palmer, N.~Pastrone, D.~Schulte and V.~Shiltsev,
  \emph{{Muon Colliders: Opening New Horizons for Particle Physics}},
  \href{https://doi.org/10.1038/s41567-020-01130-x}{\emph{Nature Phys.}
  {\bfseries 17} (2021) 289}
  [\href{https://arxiv.org/abs/2007.15684}{{\ttfamily 2007.15684}}].

\bibitem{Mohayai:2018rxn}
{\scshape MICE} collaboration, \emph{{First Demonstration of Ionization Cooling
  in MICE}},  in \emph{{Proc. 9th IPAC}}, pp.~5035--5040, JACoW Publishing,
  June, 2018, \href{https://arxiv.org/abs/1806.01807}{{\ttfamily 1806.01807}},
  \href{https://doi.org/10.18429/JACoW-IPAC2018-FRXGBE3}{DOI}.

\bibitem{Blackmore:2018mfr}
{\scshape MICE} collaboration, \emph{{Recent results from the study of
  emittance evolution in MICE}},  in \emph{{Proc. 9th IPAC}}, pp.~1699--1701,
  JACoW Publishing, June, 2018,
  \href{https://arxiv.org/abs/1806.04409}{{\ttfamily 1806.04409}},
  \href{https://doi.org/10.18429/JACoW-IPAC2018-TUPML067}{DOI}.

\bibitem{MICE:2019jkl}
{\scshape MICE} collaboration, \emph{{Demonstration of cooling by the Muon
  Ionization Cooling Experiment}},
  \href{https://doi.org/10.1038/s41586-020-1958-9}{\emph{Nature} {\bfseries
  578} (2020) 53} [\href{https://arxiv.org/abs/1907.08562}{{\ttfamily
  1907.08562}}].

\bibitem{Antonelli:2015nla}
M.~Antonelli, M.~Boscolo, R.~Di~Nardo and P.~Raimondi, \emph{{Novel proposal
  for a low emittance muon beam using positron beam on target}},
  \href{https://doi.org/10.1016/j.nima.2015.10.097}{\emph{Nucl. Instrum. Meth.
  A} {\bfseries 807} (2016) 101}
  [\href{https://arxiv.org/abs/1509.04454}{{\ttfamily 1509.04454}}].

\bibitem{Chiesa:2020awd}
M.~Chiesa, F.~Maltoni, L.~Mantani, B.~Mele, F.~Piccinini and X.~Zhao,
  \emph{{Measuring the quartic Higgs self-coupling at a multi-TeV muon
  collider}}, \href{https://doi.org/10.1007/JHEP09(2020)098}{\emph{JHEP}
  {\bfseries 2020} (2020) 098}
  [\href{https://arxiv.org/abs/2003.13628}{{\ttfamily 2003.13628}}].

\bibitem{Han:2020pif}
T.~Han, D.~Liu, I.~Low and X.~Wang, \emph{{Electroweak Couplings of the Higgs
  Boson at a Multi-TeV Muon Collider}},
  \href{https://doi.org/10.1103/PhysRevD.103.013002}{\emph{Phys. Rev. D}
  {\bfseries 103} (2021) 013002}
  [\href{https://arxiv.org/abs/2008.12204}{{\ttfamily 2008.12204}}].

\bibitem{Chiesa:2021qpr}
M.~Chiesa, B.~Mele and F.~Piccinini, \emph{{Multi Higgs production via photon
  fusion at future multi-TeV muon colliders}},
  \href{https://arxiv.org/abs/2109.10109}{{\ttfamily 2109.10109}}.

\bibitem{Cepeda:2021rql}
M.~Cepeda, S.~Gori, V.~M. Outschoorn and J.~Shelton, \emph{{Exotic Higgs
  Decays}},  \href{https://arxiv.org/abs/2111.12751}{{\ttfamily 2111.12751}}.

\bibitem{Chen:2021pqi}
J.~Chen, T.~Li, C.-T. Lu, Y.~Wu and C.-Y. Yao, \emph{{The measurement of Higgs
  self-couplings through $2\rightarrow 3$ VBS in future muon colliders}},
  \href{https://arxiv.org/abs/2112.12507}{{\ttfamily 2112.12507}}.

\bibitem{Buonincontri:2022ylv}
{\scshape Muon Collider Physics and Detector Working Group} collaboration,
  \emph{{Higgs boson couplings at muon collider}},
  \href{https://doi.org/10.22323/1.398.0619}{\emph{PoS} {\bfseries EPS-HEP2021}
  (2022) 619}.

\bibitem{Han:2020uak}
T.~Han, Z.~Liu, L.-T. Wang and X.~Wang, \emph{{WIMPs at High Energy Muon
  Colliders}}, \href{https://doi.org/10.1103/PhysRevD.103.075004}{\emph{Phys.
  Rev. D} {\bfseries 103} (2021) 075004}
  [\href{https://arxiv.org/abs/2009.11287}{{\ttfamily 2009.11287}}].

\bibitem{Capdevilla:2021fmj}
R.~Capdevilla, F.~Meloni, R.~Simoniello and J.~Zurita, \emph{{Hunting wino and
  higgsino dark matter at the muon collider with disappearing tracks}},
  \href{https://doi.org/10.1007/JHEP06(2021)133}{\emph{JHEP} {\bfseries 06}
  (2021) 133} [\href{https://arxiv.org/abs/2102.11292}{{\ttfamily
  2102.11292}}].

\bibitem{Medina:2021ram}
A.~D. Medina, N.~I. Mileo, A.~Szynkman and S.~A. Tanco, \emph{{The Elusive
  Muonic WIMP}},  \href{https://arxiv.org/abs/2112.09103}{{\ttfamily
  2112.09103}}.

\bibitem{Huang:2021biu}
G.-y. Huang, S.~Jana, F.~S. Queiroz and W.~Rodejohann, \emph{{Probing the RK(*)
  anomaly at a muon collider}},
  \href{https://doi.org/10.1103/PhysRevD.105.015013}{\emph{Phys. Rev. D}
  {\bfseries 105} (2022) 015013}
  [\href{https://arxiv.org/abs/2103.01617}{{\ttfamily 2103.01617}}].

\bibitem{Asadi:2021gah}
P.~Asadi, R.~Capdevilla, C.~Cesarotti and S.~Homiller, \emph{{Searching for
  leptoquarks at future muon colliders}},
  \href{https://doi.org/10.1007/JHEP10(2021)182}{\emph{JHEP} {\bfseries 2021}
  (2021) 182} [\href{https://arxiv.org/abs/2104.05720}{{\ttfamily
  2104.05720}}].

\bibitem{Haghighat:2021djz}
G.~Haghighat and M.~Mohammadi~Najafabadi, \emph{{Search for
  lepton-flavor-violating ALPs at a future muon collider and utilization of
  polarization-induced effects}},
  \href{https://arxiv.org/abs/2106.00505}{{\ttfamily 2106.00505}}.

\bibitem{Bandyopadhyay:2021pld}
P.~Bandyopadhyay, A.~Karan and R.~Mandal, \emph{{Distinguishing signatures of
  scalar leptoquarks at hadron and muon colliders}},
  \href{https://arxiv.org/abs/2108.06506}{{\ttfamily 2108.06506}}.

\bibitem{Capdevilla:2020qel}
R.~Capdevilla, D.~Curtin, Y.~Kahn and G.~Krnjaic, \emph{Discovering the physics
  of $(g-2)_\mu$ at future muon colliders},
  \href{https://doi.org/10.1103/PhysRevD.103.075028}{\emph{Phys. Rev. D}
  {\bfseries 103} (2021) 075028}
  [\href{https://arxiv.org/abs/2006.16277}{{\ttfamily 2006.16277}}].

\bibitem{Buttazzo:2020ibd}
D.~Buttazzo and P.~Paradisi, \emph{{Probing the muon $g-2$ anomaly with the
  Higgs boson at a muon collider}},
  \href{https://doi.org/10.1103/PhysRevD.104.075021}{\emph{Phys. Rev. D}
  {\bfseries 104} (2021) 075021}
  [\href{https://arxiv.org/abs/2012.02769}{{\ttfamily 2012.02769}}].

\bibitem{Yin:2020afe}
W.~Yin and M.~Yamaguchi, \emph{{Muon $g-2$ at multi-TeV muon collider}},
  \href{https://arxiv.org/abs/2012.03928}{{\ttfamily 2012.03928}}.

\bibitem{Capdevilla:2021rwo}
R.~Capdevilla, D.~Curtin, Y.~Kahn and G.~Krnjaic, \emph{{No-lose theorem for
  discovering the new physics of (g-2)\ensuremath{\mu} at muon colliders}},
  \href{https://doi.org/10.1103/PhysRevD.105.015028}{\emph{Phys. Rev. D}
  {\bfseries 105} (2022) 015028}
  [\href{https://arxiv.org/abs/2101.10334}{{\ttfamily 2101.10334}}].

\bibitem{Chen:2021rnl}
N.~Chen, B.~Wang and C.-Y. Yao, \emph{{The collider tests of a leptophilic
  scalar for the anomalous magnetic moments}},
  \href{https://arxiv.org/abs/2102.05619}{{\ttfamily 2102.05619}}.

\bibitem{Li:2021lnz}
T.~Li, M.~A. Schmidt, C.-Y. Yao and M.~Yuan, \emph{{Charged lepton flavor
  violation in light of the muon magnetic moment anomaly and colliders}},
  \href{https://doi.org/10.1140/epjc/s10052-021-09569-9}{\emph{Eur. Phys. J. C}
  {\bfseries 81} (2021) 811}
  [\href{https://arxiv.org/abs/2104.04494}{{\ttfamily 2104.04494}}].

\bibitem{Dermisek:2021mhi}
R.~Dermisek, K.~Hermanek and N.~McGinnis, \emph{{Di-Higgs and tri-Higgs boson
  signals of muon g-2 at a muon collider}},
  \href{https://doi.org/10.1103/PhysRevD.104.L091301}{\emph{Phys. Rev. D}
  {\bfseries 104} (2021) L091301}
  [\href{https://arxiv.org/abs/2108.10950}{{\ttfamily 2108.10950}}].

\bibitem{Capdevilla:2021kcf}
R.~Capdevilla, D.~Curtin, Y.~Kahn and G.~Krnjaic, \emph{{Systematically Testing
  Singlet Models for $(g-2)_\mu$}},
  \href{https://arxiv.org/abs/2112.08377}{{\ttfamily 2112.08377}}.

\bibitem{DiLuzio:2018jwd}
L.~Di~Luzio, R.~Gr\"ober and G.~Panico, \emph{{Probing new electroweak states
  via precision measurements at the LHC and future colliders}},
  \href{https://doi.org/10.1007/JHEP01(2019)011}{\emph{JHEP} {\bfseries 2019}
  (2019) 011} [\href{https://arxiv.org/abs/1810.10993}{{\ttfamily
  1810.10993}}].

\bibitem{Buttazzo:2020uzc}
D.~Buttazzo, R.~Franceschini and A.~Wulzer, \emph{{Two Paths Towards Precision
  at a Very High Energy Lepton Collider}},
  \href{https://doi.org/10.1007/JHEP05(2021)219}{\emph{JHEP} {\bfseries 2021}
  (2021) 219} [\href{https://arxiv.org/abs/2012.11555}{{\ttfamily
  2012.11555}}].

\bibitem{Spor:2022mxl}
S.~Spor and M.~K\"oksal, \emph{{Investigation of anomalous triple gauge
  couplings in $\mu\gamma$ collision at multi-TeV muon colliders}},
  \href{https://arxiv.org/abs/2201.00787}{{\ttfamily 2201.00787}}.

\bibitem{Chen:2022msz}
S.~Chen, A.~Glioti, R.~Rattazzi, L.~Ricci and A.~Wulzer, \emph{{Learning from
  Radiation at a Very High Energy Lepton Collider}},
  \href{https://arxiv.org/abs/2202.10509}{{\ttfamily 2202.10509}}.

\bibitem{Eichten:2013ckl}
E.~Eichten and A.~Martin, \emph{{The Muon Collider as a $H/A$ Factory}},
  \href{https://doi.org/10.1016/j.physletb.2013.11.035}{\emph{Phys. Lett. B}
  {\bfseries 728} (2014) 125}
  [\href{https://arxiv.org/abs/1306.2609}{{\ttfamily 1306.2609}}].

\bibitem{Chakrabarty:2014pja}
N.~Chakrabarty, T.~Han, Z.~Liu and B.~Mukhopadhyaya, \emph{{Radiative Return
  for Heavy Higgs Boson at a Muon Collider}},
  \href{https://doi.org/10.1103/PhysRevD.91.015008}{\emph{Phys. Rev. D}
  {\bfseries 91} (2015) 015008}
  [\href{https://arxiv.org/abs/1408.5912}{{\ttfamily 1408.5912}}].

\bibitem{Buttazzo:2018qqp}
D.~Buttazzo, D.~Redigolo, F.~Sala and A.~Tesi, \emph{{Fusing Vectors into
  Scalars at High Energy Lepton Colliders}},
  \href{https://doi.org/10.1007/JHEP11(2018)144}{\emph{JHEP} {\bfseries 2018}
  (2018) 144} [\href{https://arxiv.org/abs/1807.04743}{{\ttfamily
  1807.04743}}].

\bibitem{Bandyopadhyay:2020otm}
P.~Bandyopadhyay and A.~Costantini, \emph{{Obscure Higgs boson at Colliders}},
  \href{https://doi.org/10.1103/PhysRevD.103.015025}{\emph{Phys. Rev. D}
  {\bfseries 103} (2021) 015025}
  [\href{https://arxiv.org/abs/2010.02597}{{\ttfamily 2010.02597}}].

\bibitem{Liu:2021jyc}
W.~Liu and K.-P. Xie, \emph{{Probing electroweak phase transition with
  multi-TeV muon colliders and gravitational waves}},
  \href{https://doi.org/10.1007/JHEP04(2021)015}{\emph{JHEP} {\bfseries 2021}
  (2021) 015} [\href{https://arxiv.org/abs/2101.10469}{{\ttfamily
  2101.10469}}].

\bibitem{Han:2021udl}
T.~Han, S.~Li, S.~Su, W.~Su and Y.~Wu, \emph{{Heavy Higgs bosons in 2HDM at a
  muon collider}},
  \href{https://doi.org/10.1103/PhysRevD.104.055029}{\emph{Phys. Rev. D}
  {\bfseries 104} (2021) 055029}
  [\href{https://arxiv.org/abs/2102.08386}{{\ttfamily 2102.08386}}].

\bibitem{Sen:2021fha}
C.~Sen, P.~Bandyopadhyay, S.~Dutta and A.~KT, \emph{{Displaced Higgs production
  in Type-III Seesaw at the LHC/FCC, MATHUSLA and Muon collider}},
  \href{https://arxiv.org/abs/2107.12442}{{\ttfamily 2107.12442}}.

\bibitem{Liu:2021akf}
W.~Liu, K.-P. Xie and Z.~Yi, \emph{{Testing leptogenesis at the LHC and future
  muon colliders: a $Z'$ scenario}},
  \href{https://arxiv.org/abs/2109.15087}{{\ttfamily 2109.15087}}.

\bibitem{Cesarotti:2022ttv}
C.~Cesarotti, S.~Homiller, R.~K. Mishra and M.~Reece, \emph{{Probing New Gauge
  Forces with a High-Energy Muon Beam Dump}},
  \href{https://arxiv.org/abs/2202.12302}{{\ttfamily 2202.12302}}.

\bibitem{Bandyopadhyay:2020mnp}
P.~Bandyopadhyay, A.~Karan and C.~Sen, \emph{{Discerning Signatures of Seesaw
  Models and Complementarity of Leptonic Colliders}},
  \href{https://arxiv.org/abs/2011.04191}{{\ttfamily 2011.04191}}.

\bibitem{Costantini:2020stv}
A.~Costantini, F.~De~Lillo, F.~Maltoni, L.~Mantani, O.~Mattelaer, R.~Ruiz
  et~al., \emph{{Vector boson fusion at multi-TeV muon colliders}},
  \href{https://doi.org/10.1007/JHEP09(2020)080}{\emph{JHEP} {\bfseries 2020}
  (2020) 080} [\href{https://arxiv.org/abs/2005.10289}{{\ttfamily
  2005.10289}}].

\bibitem{AlAli:2021let}
H.~Al~Ali et~al., \emph{{The Muon Smasher's Guide}},
  \href{https://arxiv.org/abs/2103.14043}{{\ttfamily 2103.14043}}.

\bibitem{Franceschini:2021aqd}
R.~Franceschini and M.~Greco, \emph{{Higgs and BSM Physics at the Future Muon
  Collider}}, \href{https://doi.org/10.3390/sym13050851}{\emph{Symmetry}
  {\bfseries 13} (2021) 851}
  [\href{https://arxiv.org/abs/2104.05770}{{\ttfamily 2104.05770}}].

\bibitem{Brivio2017}
I.~Brivio, M.~B. Gavela, L.~Merlo, K.~Mimasu, J.~M. No, R.~del Rey et~al.,
  \emph{{ALPs} effective field theory and collider signatures},
  \href{https://doi.org/10.1140/epjc/s10052-017-5111-3}{\emph{Eur. Phys. J. C}
  {\bfseries 77} (2017) 572}
  [\href{https://arxiv.org/abs/1701.05379}{{\ttfamily 1701.05379}}].

\bibitem{Alonso-Alvarez:2018irt}
G.~Alonso-\'Alvarez, M.~B. Gavela and P.~Quilez, \emph{Axion couplings to
  electroweak gauge bosons},
  \href{https://doi.org/10.1140/epjc/s10052-019-6732-5}{\emph{Eur. Phys. J. C}
  {\bfseries 79} (2019) 223}
  [\href{https://arxiv.org/abs/1811.05466}{{\ttfamily 1811.05466}}].

\bibitem{BlindToAnomaly}
J.~Quevillon and C.~Smith, \emph{Axions are blind to anomalies},
  \href{https://doi.org/10.1140/epjc/s10052-019-7304-4}{\emph{Eur. Phys. J. C}
  {\bfseries 79} (2019) 822}
  [\href{https://arxiv.org/abs/1903.12559}{{\ttfamily 1903.12559}}].

\bibitem{Bonnefoy2020}
Q.~Bonnefoy, L.~D. Luzio, C.~Grojean, A.~Paul and A.~N. Rossia, \emph{The
  anomalous case of axion {EFT}s and massive chiral gauge fields},
  \href{https://doi.org/10.1007/JHEP07(2021)189}{\emph{JHEP} {\bfseries 2021}
  (2021) 189} [\href{https://arxiv.org/abs/2011.10025}{{\ttfamily
  2011.10025}}].

\bibitem{Quevillon:2021sfz}
J.~Quevillon, C.~Smith and P.~N.~H. Vuong, \emph{{Axion Effective Action}},
  \href{https://arxiv.org/abs/2112.00553}{{\ttfamily 2112.00553}}.

\bibitem{Bonilla:2021ufe}
J.~Bonilla, I.~Brivio, M.~B. Gavela and V.~Sanz, \emph{{One-loop corrections to
  ALPs couplings}}, \href{https://doi.org/10.1007/JHEP11(2021)168}{\emph{JHEP}
  {\bfseries 2021} (2021) 168}
  [\href{https://arxiv.org/abs/2107.11392}{{\ttfamily 2107.11392}}].

\bibitem{Srednicki1985}
M.~Srednicki, \emph{Axion couplings to matter},
  \href{https://doi.org/10.1016/0550-3213(85)90054-9}{\emph{Nuclear Phys. B
  Proc. Suppl.} {\bfseries 260} (1985) 689}.

\bibitem{Buen-Abad2021}
M.~A. Buen-Abad, J.~Fan, M.~Reece and C.~Sun, \emph{{Challenges for an axion
  explanation of the muon $g-2$ measurement}},
  \href{https://doi.org/10.1007/JHEP09(2021)101}{\emph{JHEP} {\bfseries 2021}
  (2021) 101} [\href{https://arxiv.org/abs/2104.03267}{{\ttfamily
  2104.03267}}].

\bibitem{CMS:2016kgr}
{\scshape CMS} collaboration, \emph{{Search for high-mass diphoton resonances
  in proton\textendash{}proton collisions at 13 TeV and combination with 8 TeV
  search}}, \href{https://doi.org/10.1016/j.physletb.2017.01.027}{\emph{Phys.
  Lett. B} {\bfseries 767} (2017) 147}
  [\href{https://arxiv.org/abs/1609.02507}{{\ttfamily 1609.02507}}].

\bibitem{ATLAS:2021uiz}
{\scshape ATLAS} collaboration, \emph{{Search for resonances decaying into
  photon pairs in 139 fb$^{-1}$ of $pp$ collisions at $\sqrt{s} =$ 13 TeV with
  the ATLAS detector}},
  \href{https://doi.org/10.1016/j.physletb.2021.136651}{\emph{Phys. Lett. B}
  {\bfseries 822} (2021) 136651}
  [\href{https://arxiv.org/abs/2102.13405}{{\ttfamily 2102.13405}}].

\bibitem{ATLAS:2018sxj}
{\scshape ATLAS} collaboration, \emph{{Search for heavy resonances decaying to
  a photon and a hadronically decaying $Z/W/H$ boson in $pp$ collisions at
  $\sqrt{s}=13$ $\mathrm{TeV}$ with the ATLAS detector}},
  \href{https://doi.org/10.1103/PhysRevD.98.032015}{\emph{Phys. Rev. D}
  {\bfseries 98} (2018) 032015}
  [\href{https://arxiv.org/abs/1805.01908}{{\ttfamily 1805.01908}}].

\bibitem{ATLAS:2018sbw}
{\scshape ATLAS} collaboration, \emph{{Combination of searches for heavy
  resonances decaying into bosonic and leptonic final states using 36 fb$^{-1}$
  of proton-proton collision data at $\sqrt{s} = 13$ TeV with the ATLAS
  detector}}, \href{https://doi.org/10.1103/PhysRevD.98.052008}{\emph{Phys.
  Rev. D} {\bfseries 98} (2018) 052008}
  [\href{https://arxiv.org/abs/1808.02380}{{\ttfamily 1808.02380}}].

\bibitem{CMS:2019qem}
{\scshape CMS} collaboration, \emph{{A multi-dimensional search for new heavy
  resonances decaying to boosted WW, WZ, or ZZ boson pairs in the dijet final
  state at 13 TeV}},
  \href{https://doi.org/10.1140/epjc/s10052-020-7773-5}{\emph{Eur. Phys. J. C}
  {\bfseries 80} (2020) 237}
  [\href{https://arxiv.org/abs/1906.05977}{{\ttfamily 1906.05977}}].

\bibitem{ATLAS:2020fry}
{\scshape ATLAS} collaboration, \emph{{Search for heavy diboson resonances in
  semileptonic final states in pp collisions at $\sqrt{s}=13$ TeV with the
  ATLAS detector}},
  \href{https://doi.org/10.1140/epjc/s10052-020-08554-y}{\emph{Eur. Phys. J. C}
  {\bfseries 80} (2020) 1165}
  [\href{https://arxiv.org/abs/2004.14636}{{\ttfamily 2004.14636}}].

\bibitem{ATLAS:2016mti}
{\scshape ATLAS} collaboration, \emph{{Search for heavy resonances decaying to
  a $Z$ boson and a photon in $pp$ collisions at $\sqrt{s}=13$~TeV with the
  ATLAS detector}},
  \href{https://doi.org/10.1016/j.physletb.2016.11.005}{\emph{Phys. Lett. B}
  {\bfseries 764} (2017) 11}
  [\href{https://arxiv.org/abs/1607.06363}{{\ttfamily 1607.06363}}].

\bibitem{CMS:2017dyb}
{\scshape CMS} collaboration, \emph{{Search for Z$\gamma$ resonances using
  leptonic and hadronic final states in proton-proton collisions at $\sqrt{s}=$
  13 TeV}}, \href{https://doi.org/10.1007/JHEP09(2018)148}{\emph{JHEP}
  {\bfseries 2018} (2018) 148}
  [\href{https://arxiv.org/abs/1712.03143}{{\ttfamily 1712.03143}}].

\bibitem{Cacciari:2008gn}
M.~Cacciari, G.~P. Salam and G.~Soyez, \emph{The catchment area of jets},
  \href{https://doi.org/10.1088/1126-6708/2008/04/005}{\emph{JHEP} {\bfseries
  2008} (2008) 005} [\href{https://arxiv.org/abs/0802.1188}{{\ttfamily
  0802.1188}}].

\bibitem{Molinaro:2017rpe}
E.~Molinaro and N.~Vignaroli, \emph{{Diphoton Resonances at the LHC}},
  \href{https://doi.org/10.1142/S0217732317300245}{\emph{Mod. Phys. Lett. A}
  {\bfseries 32} (2017) 1730024}
  [\href{https://arxiv.org/abs/1707.00926}{{\ttfamily 1707.00926}}].

\bibitem{Manohar:2016nzj}
A.~Manohar, P.~Nason, G.~P. Salam and G.~Zanderighi, \emph{{How bright is the
  proton? A precise determination of the photon parton distribution function}},
  \href{https://doi.org/10.1103/PhysRevLett.117.242002}{\emph{Phys. Rev. Lett.}
  {\bfseries 117} (2016) 242002}
  [\href{https://arxiv.org/abs/1607.04266}{{\ttfamily 1607.04266}}].

\bibitem{Fornal:2018znf}
B.~Fornal, A.~V. Manohar and W.~J. Waalewijn, \emph{{Electroweak Gauge Boson
  Parton Distribution Functions}},
  \href{https://doi.org/10.1007/JHEP05(2018)106}{\emph{JHEP} {\bfseries 05}
  (2018) 106} [\href{https://arxiv.org/abs/1803.06347}{{\ttfamily
  1803.06347}}].

\bibitem{Buarque:2021dji}
D.~Buarque et~al., \emph{{Vector Boson Scattering Processes: Status and
  Prospects}},  \href{https://arxiv.org/abs/2106.01393}{{\ttfamily
  2106.01393}}.

\bibitem{CMS:2013xfa}
{\scshape CMS} collaboration, \emph{{Projected Performance of an Upgraded CMS
  Detector at the LHC and HL-LHC: Contribution to the Snowmass Process}},  in
  \emph{{CSS2013}}, July, 2013,
  \href{https://arxiv.org/abs/1307.7135}{{\ttfamily 1307.7135}}.

\bibitem{Butler:2019rpu}
{\scshape CMS} collaboration, \emph{{A MIP Timing Detector for the CMS Phase-2
  Upgrade}}, .

\bibitem{Komiske:2017ubm}
P.~T. Komiske, E.~M. Metodiev, B.~Nachman and M.~D. Schwartz, \emph{{Pileup
  Mitigation with Machine Learning (PUMML)}},
  \href{https://doi.org/10.1007/JHEP12(2017)051}{\emph{JHEP} {\bfseries 2017}
  (2017) 051} [\href{https://arxiv.org/abs/1707.08600}{{\ttfamily
  1707.08600}}].

\bibitem{ArjonaMartinez:2018eah}
J.~Arjona~Mart\'\i{}nez, O.~Cerri, M.~Pierini, M.~Spiropulu and J.-R. Vlimant,
  \emph{{Pileup mitigation at the Large Hadron Collider with graph neural
  networks}}, \href{https://doi.org/10.1140/epjp/i2019-12710-3}{\emph{Eur.
  Phys. J. Plus} {\bfseries 134} (2019) 333}
  [\href{https://arxiv.org/abs/1810.07988}{{\ttfamily 1810.07988}}].

\bibitem{CMS:2021jji}
{\scshape CMS} collaboration, \emph{{Measurements of the pp $\to$
  W$^\pm\gamma\gamma$ and pp $\to$ Z$\gamma\gamma$ cross sections at $\sqrt{s}
  =$ 13 TeV and limits on anomalous quartic gauge couplings}},
  \href{https://doi.org/10.1007/JHEP10(2021)174}{\emph{JHEP} {\bfseries 2021}
  (2021) 174} [\href{https://arxiv.org/abs/2105.12780}{{\ttfamily
  2105.12780}}].

\bibitem{ATLAS:2019dny}
{\scshape ATLAS} collaboration, \emph{{Evidence for the production of three
  massive vector bosons with the ATLAS detector}},
  \href{https://doi.org/10.1016/j.physletb.2019.134913}{\emph{Phys. Lett. B}
  {\bfseries 798} (2019) 134913}
  [\href{https://arxiv.org/abs/1903.10415}{{\ttfamily 1903.10415}}].

\bibitem{CMS:2020hjs}
{\scshape CMS} collaboration, \emph{{Observation of the Production of Three
  Massive Gauge Bosons at $\sqrt {s}$ =13 TeV}},
  \href{https://doi.org/10.1103/PhysRevLett.125.151802}{\emph{Phys. Rev. Lett.}
  {\bfseries 125} (2020) 151802}
  [\href{https://arxiv.org/abs/2006.11191}{{\ttfamily 2006.11191}}].

\bibitem{CMS:2019mpq}
{\scshape CMS} collaboration, \emph{{Search for the production of
  W$^\pm$W$^\pm$W$^\mp$ events at $\sqrt{s} =$ 13 TeV}},
  \href{https://doi.org/10.1103/PhysRevD.100.012004}{\emph{Phys. Rev. D}
  {\bfseries 100} (2019) 012004}
  [\href{https://arxiv.org/abs/1905.04246}{{\ttfamily 1905.04246}}].

\bibitem{ATLAS:2017bon}
{\scshape ATLAS} collaboration, \emph{{Study of $WW\gamma$ and $WZ\gamma$
  production in $pp$ collisions at $\sqrt{s} = 8$ TeV and search for anomalous
  quartic gauge couplings with the ATLAS experiment}},
  \href{https://doi.org/10.1140/epjc/s10052-017-5180-3}{\emph{Eur. Phys. J. C}
  {\bfseries 77} (2017) 646}
  [\href{https://arxiv.org/abs/1707.05597}{{\ttfamily 1707.05597}}].

\bibitem{Bonilla:2022pxu}
J.~Bonilla, I.~Brivio, J.~Machado-Rodr\'\i{}guez and J.~F. de~Troc\'oniz,
  \emph{{Nonresonant Searches for Axion-Like Particles in Vector Boson
  Scattering Processes at the LHC}},
  \href{https://arxiv.org/abs/2202.03450}{{\ttfamily 2202.03450}}.

\bibitem{Green:2016trm}
D.~R. Green, P.~Meade and M.-A. Pleier, \emph{{Multiboson interactions at the
  LHC}}, \href{https://doi.org/10.1103/RevModPhys.89.035008}{\emph{Rev. Mod.
  Phys.} {\bfseries 89} (2017) 035008}
  [\href{https://arxiv.org/abs/1610.07572}{{\ttfamily 1610.07572}}].

\bibitem{Azzi:2019yne}
P.~Azzi et~al., \emph{{Report from Working Group 1}: {Standard Model Physics at
  the HL-LHC and HE-LHC}},
  \href{https://doi.org/10.23731/CYRM-2019-007.1}{\emph{CERN Yellow Rep.
  Monogr.} {\bfseries 7} (2019) 1}
  [\href{https://arxiv.org/abs/1902.04070}{{\ttfamily 1902.04070}}].

\bibitem{Alwall:2014hca}
J.~Alwall, R.~Frederix, S.~Frixione, V.~Hirschi, F.~Maltoni, O.~Mattelaer
  et~al., \emph{{The automated computation of tree-level and next-to-leading
  order differential cross sections, and their matching to parton shower
  simulations}}, \href{https://doi.org/10.1007/JHEP07(2014)079}{\emph{JHEP}
  {\bfseries 2014} (2014) 079}
  [\href{https://arxiv.org/abs/1405.0301}{{\ttfamily 1405.0301}}].

\bibitem{Sjostrand:2007gs}
T.~Sjostrand, S.~Mrenna and P.~Z. Skands, \emph{A brief introduction to
  {PYTHIA} 8.1}, \href{https://doi.org/10.1016/j.cpc.2008.01.036}{\emph{Comput.
  Phys. Commun.} {\bfseries 178} (2008) 852}
  [\href{https://arxiv.org/abs/0710.3820}{{\ttfamily 0710.3820}}].

\bibitem{deFavereau:2013fsa}
{\scshape DELPHES 3} collaboration, \emph{{DELPHES} 3: a modular framework for
  fast simulation of a generic collider experiment},
  \href{https://doi.org/10.1007/JHEP02(2014)057}{\emph{JHEP} {\bfseries 2014}
  (2014) 057} [\href{https://arxiv.org/abs/1307.6346}{{\ttfamily 1307.6346}}].

\bibitem{Boronat:2014hva}
M.~Boronat, J.~Fuster, I.~Garcia, E.~Ros and M.~Vos, \emph{{A robust jet
  reconstruction algorithm for high-energy lepton colliders}},
  \href{https://doi.org/10.1016/j.physletb.2015.08.055}{\emph{Phys. Lett. B}
  {\bfseries 750} (2015) 95} [\href{https://arxiv.org/abs/1404.4294}{{\ttfamily
  1404.4294}}].

\bibitem{vonWeizsacker:1934nji}
C.~F. von Weizsacker, \emph{{Radiation emitted in collisions of very fast
  electrons}}, \href{https://doi.org/10.1007/BF01333110}{\emph{Z. Phys.}
  {\bfseries 88} (1934) 612}.

\bibitem{Williams:1934ad}
E.~J. Williams, \emph{{Nature of the high-energy particles of penetrating
  radiation and status of ionization and radiation formulae}},
  \href{https://doi.org/10.1103/PhysRev.45.729}{\emph{Phys. Rev.} {\bfseries
  45} (1934) 729}.

\bibitem{Ruiz:2021tdt}
R.~Ruiz, A.~Costantini, F.~Maltoni and O.~Mattelaer, \emph{{The Effective
  Vector Boson Approximation in High-Energy Muon Collisions}},
  \href{https://arxiv.org/abs/2111.02442}{{\ttfamily 2111.02442}}.

\bibitem{Cowan:2010js}
G.~Cowan, K.~Cranmer, E.~Gross and O.~Vitells, \emph{{Asymptotic formulae for
  likelihood-based tests of new physics}},
  \href{https://doi.org/10.1140/epjc/s10052-011-1554-0}{\emph{Eur. Phys. J. C}
  {\bfseries 71} (2011) 1554}
  [\href{https://arxiv.org/abs/1007.1727}{{\ttfamily 1007.1727}}].

\bibitem{Ren:2014bza}
H.-Y. Ren, L.-H. Xia and Y.-P. Kuang, \emph{Model-independent probe of
  anomalous heavy neutral higgs bosons at the {LHC}},
  \href{https://doi.org/10.1103/PhysRevD.90.115002}{\emph{Phys. Rev. D}
  {\bfseries 90} (2014) 115002}
  [\href{https://arxiv.org/abs/1404.6367}{{\ttfamily 1404.6367}}].

\bibitem{Antoniadis:2009ze}
I.~Antoniadis, A.~Boyarsky, S.~Espahbodi, O.~Ruchayskiy and J.~D. Wells,
  \emph{{Anomaly driven signatures of new invisible physics at the Large Hadron
  Collider}},
  \href{https://doi.org/10.1016/j.nuclphysb.2009.09.009}{\emph{Nucl. Phys. B}
  {\bfseries 824} (2010) 296}
  [\href{https://arxiv.org/abs/0901.0639}{{\ttfamily 0901.0639}}].

\bibitem{Bondarenko:2019tss}
K.~Bondarenko, A.~Boyarsky, M.~Ovchynnikov, O.~Ruchayskiy and L.~Shchutska,
  \emph{{Probing new physics with displaced vertices: muon tracker at CMS}},
  \href{https://doi.org/10.1103/PhysRevD.100.075015}{\emph{Phys. Rev. D}
  {\bfseries 100} (2019) 075015}
  [\href{https://arxiv.org/abs/1903.11918}{{\ttfamily 1903.11918}}].

\bibitem{Han:2022mzp}
T.~Han, T.~Li and X.~Wang, \emph{{Axion-Like Particles at High Energy Muon
  Colliders -- A White paper for Snowmass 2021}},  in \emph{{2022 Snowmass
  Summer Study}}, 3, 2022, \href{https://arxiv.org/abs/2203.05484}{{\ttfamily
  2203.05484}}.

\end{thebibliography}\endgroup
\end{document}